\documentclass[11pt]{article}

\usepackage{amsmath}
\usepackage{enumitem}
\usepackage{multirow}
\usepackage{hhline}
\usepackage{bm}
\usepackage{setspace}
\usepackage{graphicx}
\usepackage{dsfont}
\usepackage{epstopdf}
\usepackage{hyperref}
\usepackage{caption}
\usepackage{subcaption}
\usepackage{pdflscape}
\usepackage{framed}
\usepackage{color}
\usepackage{braket}

\usepackage{multirow}
\usepackage{amssymb, graphics}
\usepackage[affil-it]{authblk}
\usepackage{lipsum}
\usepackage{titlesec}

\titlespacing\section{0pt}{12pt plus 4pt minus 2pt}{-2pt plus 2pt minus 2pt}
\titlespacing\subsection{0pt}{12pt plus 4pt minus 2pt}{-4pt plus 2pt minus 2pt}
\titlespacing\subsubsection{0pt}{12pt plus 4pt minus 2pt}{0pt plus 2pt minus 2pt}

\newcommand{\mathsym}[1]{{}}
\newcommand{\unicode}[1]{{}}

\usepackage{wrapfig}
\usepackage{hhline}
\usepackage{bm}
\usepackage{units}
\usepackage{xparse}
\usepackage{booktabs} 
\usepackage[absolute,overlay]{textpos}
\usepackage{amsmath} 
\newcommand\scalemath[2]{\scalebox{#1}{\mbox{\ensuremath{\displaystyle #2}}}}

\def\({\left(}
\def\){\right)}
\def\[{\left[}
\def\]{\right]}

\def\n{\\ \noindent \newline}
\definecolor{darkorange}{rgb}{1.0, 0.55, 0.0}

\begin{document}
\singlespacing
\onehalfspacing

\renewcommand{\thefootnote}{\fnsymbol{footnote}}

\thispagestyle{empty}
\vspace{-.5in}
\begin{flushright}
OSU-HEP-15-04 \\
CETUP2015-014
\end{flushright}

\vspace*{1.0cm}
\begin{center}
\LARGE {\bf \Large A Minimal Non-Supersymmetric $SO(10)$ Model: Gauge Coupling Unification, Proton Decay and Fermion masses}
\end{center}

\begin{center}

{\large \bf K.S. Babu$^a$\footnote{Email:
babu@okstate.edu} and S. Khan$^a$\footnote{Email:
saki.khan@okstate.edu}}\n
\end{center}

\vspace{-0.7in}
\begin{center}
\it $^a $Department of Physics, Oklahoma State University, Stillwater, Oklahoma 74078, USA
\end{center}
\vspace{.5in}
\renewcommand{\thefootnote}{\arabic{footnote}}
\setcounter{footnote}{0}

\begin{abstract}
We present a minimal renormalizable non-supersymmetric $SO(10)$ grand unified model with a symmetry breaking sector consisting of Higgs fields in the $54_H + 126_H + 10_H$ representations. This model admits a single intermediate scale associated with Pati-Salam symmetry along with a discrete parity. Spontaneous symmetry breaking, the unification of gauge couplings and proton lifetime estimates are studied in detail in this framework. Including threshold corrections self-consistently, obtained from a full analysis of the Higgs potential, we show that the model is compatible with the current experimental bound on proton lifetime. The model generally predicts an upper bound of few times $10^{35}\; \text{yrs}$ for proton lifetime, which is not too far from the present Super-Kamiokande limit of $\tau_p \gtrsim 1.29 \times 10^{34}\; \text{yrs}$. With the help of a Pecci-Quinn symmetry and the resulting axion, the model provides a suitable dark matter candidate while also solving the strong CP problem. The intermediate scale, $M_I \approx (10^{13}-10^{14})\;\text{GeV}$ which is also the $B-L$ scale, is of the right order for the right-handed neutrino mass which enables a successful description of light neutrino masses and oscillations. The Yukawa sector of the model consists of only two matrices in family space and leads to a predictive scenario for quark and lepton masses and mixings. The branching ratios for proton decay are calculable with the leading modes being $p \rightarrow e^+ \pi^0$ and $p \rightarrow \overline{\nu} \pi^+$. Even though the model predicts no new physics within the reach of LHC, the next generation proton decay detectors and axion search experiments have the capability to pass verdict on this minimal scenario.
\end{abstract}
\newpage
\section{Introduction}
The desire to achieve true unification of the strong, weak and electromagnetic forces under one simple non-abelian gauge group gave birth to the idea of Grand Unified Theories (GUTs) \cite{Pati:1973rp, Pati:1974yy, Georgi:1974sy, Georgi:1974yf}. The absence of an abelian factor in such unified theories readily quantizes electric charges, an observational feature left unexplained in the Standard Model (SM), which has served as one of the primary motivations of GUTs \cite{Georgi:1974sy}. Yet, the initially introduced minimal $SU(5)$ model fails to unify the three gauge couplings \cite{susysu5}. The $SU(3)$-color and $SU(2)$-weak gauge couplings meet around $10^{16}\; \text{GeV}$ while the $U(1)$-hypercharge gauge coupling meets $SU(2)$ gauge coupling at a much lower energy scale of $10^{13}\; \text{GeV}$, which is too low to comply with the experimental limits on proton lifetime. Of course, such an issue is absent in a low energy supersymmtric (SUSY) $SU(5)$ GUT where the three gauge couplings merge to a common value around $2 \times 10^{16}\; \text{GeV}$\footnote{In supersymmetry GUTs, the rate for proton decay $\Gamma(p \rightarrow \overline{\nu} K^+)$ arising from color triplet Higgsino exchange has to comply with experimental limits, which is a non-trivial task for SUSY GUT model building.} \cite{susysu5}.

The discovery of a Standard Model-like Higgs boson became the crowning event of the first run of LHC \cite{higgsdiscovery}. The triumph of SM and the absence of any compelling evidence (such as signals for new particles or exotic phenomena) of physics beyond SM in the LHC data, are making a large portion of the physics community rethink about the next step in the field. Supersymmetry (SUSY), one of the most elegant and successful solutions of the hierarchy problem \cite{hierarchyproblem} along with the WIMP (Weakly Interacting Massive Particle) scenario for dark matter, has been the most widely studied candidate of physics beyond SM at the TeV scale. Even though there is absolutely no reason to abandon the hopes of finding the necessary traces of new physics to solve such issues in the second run of LHC, one should also entertain the possibility that the hierarchy problem may simply be ``solved" by fine-tuning. There exists a variety of approaches to justify such fine-tuning \cite{Bajc:2005zf, Altarelli:2013aqa}. One might invoke the anthropic principle, which has been much talked about in the field of cosmology \cite{anthropicprinciple} as the observed value of the cosmological constant $\Lambda$ poses an unsolved naturalness problem of larger magnitude \cite{Frieman:2008sn}. If one considers a universe with domains which can have different values of some of the underlying parameters like the Higgs boson mass, it has been argued that the observed values of the masses are reasonably typical of the anthropically allowed ranges \cite{Agrawal:1997gf}. One can look for answers in the much more debatable idea of infinite number of universes (multiverse) continuously created by quantum fluctuations and we happen to live in a very unlikely one \cite{Guth:2007ng}.  In string theory landscape picture, our universe might be just one example out of $~10^{500}$ possible solutions  \cite{stringlandscaperev}. Or the ``hierarchy problem" may very well be an artificially created one in quantum field theories which necessarily require regularization of infinities. Despite the philosophical point of view one might adopt, the lack of hard evidence of new physics sparked the revival of a class of BSMs (Beyond Standard Models) known as non-SUSY Grand Unified Theories, which ignores the hierarchy problem while trying to remain consistent with all the phenomenological constraints and predicting their own experimental signatures \cite{Bajc:2005zf,quantumvacua45,Bertolinimalinskyluzio,Bertolini:2013vta, Joshipura:2011nn,Buccella:2012kc,Altarelli:2013aqa, Mambrini:2015vna}.

$SO(10)$ grand unified theory \cite{SO(10):Introduced} is undoubtedly the best motivated candidate in the above-mentioned class of models. Instead of taking help from supersymmetry  to unify all the three gauge couplings, it relies on the fact that the rank-5 $SO(10)$ group can accommodate one or more intermediate scales between the unified scale and the weak scale \cite{GoranIntSc:1981,Caswell:1982fx, Gipson:1984aj,Chang:1984qr,Deshpande:1992au,Deshpande:1992em,Bertolini:2009qj}. As the gauge group structure changes (for example, $U(1)$ is usually embedded in $SU(2)_R$) above the intermediate scale, so does the running of the gauge couplings, allowing for the possibility of unification of all three gauge couplings. The fact that $SO(10)$ GUT is naturally free of anomalies  \cite{SO(10):Introduced} and that it unifies one generation of fermions (both leptons and quarks) into a 16-dimensional spinorial representation $(16_F)$ only enhance the beauty of the model. This is due to the fact that the $SO(10)$ symmetry includes quark-lepton symmetry ($SU(4)_C$) \cite{Pati:1974yy} and the left-right symmetry ($SU(2)_L \times SU(2)_R$) \cite{LRmodel}. Unlike $SU(5)$, the $16_F$ representation also includes a right-handed neutrino and provides an appealing explanation of small neutrino masses and oscillations through the seesaw mechanism \cite{seesawmodel, SchechterValle}. This setup has all the ingredients to explain the observed baryon asymmetry of the universe either by leptogenesis \cite{Fukugita:1986hr, Davidson:2008bu} or by $B-L$ violating decays of new scalar states \cite{Babu:2012Postspheleron, Babu:2012iv}.

Our goal in this paper is to construct the most minimal non-supersymmetric $SO(10)$ model. We shall be guided by simplicity and minimality, while being consistent with proton lifetime bound \cite{Nishino:2012ipa, Babu:2013jba}, staying in agreement with the fermion masses and mixings \cite{Joshipura:2011nn}, providing axion as a suitable candidate for dark matter while solving the strong CP problem \cite{Peccei:1977hh}. The model should be able to address the issue of the instability of the electroweak vacuum\footnote {The study of the stability of the SM eletroweak vacuum has shown that for a Higgs mass of $125.5\pm0.5\; \text{GeV}$ the Higgs quartic coupling of SM becomes negative around $(10^{10}-10^{11})\;\text{GeV}$ energy scale \cite{electroweakstability} , indicating that we might be living in a metastable universe. The electroweak vacuum can be stabilized by the threshold effect of a single scalar with vacuum expectation value close to the instability energy scale \cite{EliasMiro:2012ay, Salvio:2015cja}. Such a scalar arises naturally in our framework as a remnant of the PQ symmetry breaking.} with SM singlet(s) and other particles lying inside the Higgs sector of the model. Inflation might be generated by a gravitationally coupled SM singlet(s) outside the model, or SM singlet(s) already present in the model may do the trick.

Search for such a minimal, yet realistic, unified model is highly non-trivial as the constraints provided by phenomenology are quiet demanding. After going through the process of selecting the Higgs sector, we end up with a symmetry breaking pattern:
\begin{align*}
SO(10) \xrightarrow{54_H} SU(4)_C \times SU(2)_L \times SU(2)_R \times D \xrightarrow{126_H}  SU(3)_C &\times SU(2)_L \times U(1)_Y \\
& \xrightarrow{10_H} SU(3)_C \times U(1)_{em} .
\end{align*}
From the viewpoint of minimal Higgs sector, this symmetry breaking chain is the simplest, employing a single real $54_H$, a complex $126_H$ and a complex $10_H$. Even though earlier works \cite{Deshpande:1992au,Deshpande:1992em,Bertolini:2009qj,Abud:2012xp} may have prematurely sentenced this model as an unrealistic candidate for its failure to provide a high enough energy scale of gauge coupling unification to be compatible with proton lifetime limits, we decided to analyze the model more carefully before passing out the final verdict. Our detailed analysis of the model included the threshold corrections coming from the scalar and gauge boson sectors with complete mass spectrum. We also include effects of introducing the PQ-symmetry and its breaking by a singlet scalar, requiring compatibility with realistic and predictable Yukawa sector, and fine-tuning the hierarchy issue in the Higgs doublet sector.

We find from an explicit computation of the Higgs boson masses obtained by analyzing the Higgs potential of the model for the first time, that indeed compatibility with proton lifetime can be achieved. Rather than assuming the heavy Higgs bosons to have masses spread over an order of magnitude either way from the intermediate or GUT scale that has been employed traditionally, we chose the fundamental couplings of the Higgs potential to vary within a reasonable range, which provides more stringent constraints.

The outline of the paper is as follows. In Sec. \ref{sec:The Model}, we present the model with its symmery breaking sector and using minimal fine-tune condition - also known as extended survival hypothesis - we determine the energy scale of the various Higgs multiplets. The evolution of the gauge couplings using one-loop and two-loop renormalization group equations have been discussed in the Sec. \ref{sec:Running of gauge couplings using two loop Renormalization Group Equations} while the issue of one-loop threshold corrections due to the randomness of the scalar masses has been addressed in the Sec \ref{sec:Threshold Correction}. In Sec. \ref{sec:Proton Lifetime}, we discuss the most important prediction of any GUT, namely proton decay and its lifetime. In Sec. \ref{sec:The $54_H + 126_H$ Higgs Model for $SO(10)$ symmetry breaking}, we perform a detailed analysis of the Higgs potential and the kinetic part of the Lagrangian to determine the scalar and gauge boson mass spectrum, while in Sec. \ref{sec:Yukawa sector of the model} we analyze the Yukawa sector of the Lagrangian. Sec. \ref{sec:Technical Details} is dedicated to some technical details of the procedures that have been carried out to generate the benchmark points of the model given in Sec. \ref{sec:Results with Benchmark points}. In Sec. \ref{sec: Proton branching ratio}, we study the proton decay branching ratios in the model and in Sec. \ref{sec:Axions as Dark Matter}, we study the issues of axion as the dark matter candidate in various cosmological scenarios. Finally in Sec. \ref{sec:Conclusion} we conclude.

\section{The Model}
\label{sec:The Model}
In this section we present our logic for choosing the symmetry breaking sector and describe qualitatively the emergence of an intermediate scale. After establishing the symmetry breaking pattern, we determine the energy scales associated with each Higgs multiplet in such a way that the arrangement is in  agreement with phenomenological constraints.
\subsection{Choice of the Higgs sector}
The representations of the Higgs bosons primarily dictate the breaking of any higher gauge symmetry group down to lower ones \cite{Li:1973mq}. Various low dimensional scalar representations - $10_H, 16_H, 45_H, 54_H, 120_H, 126_H, 144_H, 210_H$-plets - have been used to break the $SO(10)$ group. Depending on the choice of the Higgs, it is possible to find multiple distinct breaking chains all of which end up in the SM symmetry. In this work, we are guided by the philosophy of minimality while staying within the perimeter of phenomenological constraints. As minimal non-supersymmtric $SU(5)$ (with no light exotics) corresponds to a proton lifetime which has been already ruled out by Super-Komiakande \cite{Nishino:2012ipa}, between the two maximal subgroup of $SO(10)$ that contains SM, namely $SU(5) \times U(1)_X$ and $SU(4)_C \times SU(2)_L \times SU(2)_R$, the latter one is preferred in the breaking chain.

A simple choice of Higgs sector consisting $45_H+16_H$ tends to break $SO(10)$ to $SU(5)$. Alternative breaking channels fail to possess a gauge hierarchy at tree level in which $SU(2)_L \times SU(2)_R \times U(1)$ or $SU(4)_C$ makes an appearance resulting in a tachyonic mass spectrum \cite{Yasue:1980fy, Anastaze:1983zk, Babu:1984mz}. The same comment holds for the Higgs system consisting of $45_H+126_H$. Recently quantum salvation of these type of models has been shown by assuming that loop level contribution to the Higgs masses surpass the tree level ones \cite{Bertolini:2009es}. In a series of papers, the details of these type of models were discussed \cite{Bertolinimalinskyluzio}. While interesting, we view such models as not the very minimal, at least in the technical sense, since loop corrections are essential to stabilize the tree level potential.

The next choice is naturally $54_H+16_H$. But the absence of any non-trivial cross-coupling between $54_H$ and $16_H$ promotes the global symmetry to $SO(10) \times SO(10)$ in this case. $54_H$ breaks one of them down to Pati-Salam (PS) symmetry with D-parity\footnote{D-parity is a discrete symmetry residing in the $SO(10)$ group which behaves similar to parity $(P)$ or charge conjugation $(C)$ operator, at least in the case of fermions (for example, $D(4,1,2)_{PS}=(\overline{4},2,1)_{PS}$ or simply $q\rightarrow q^C$). For the scalar sector, the effect of D-parity is a little different from $C$ or $P$. For example, both $54_H$ and $210_H$ lead to the maximal little groups being the Pati-Salam group, as both possess a Pati-Salam singlet $(1,1,1)_{PS}$. Yet the singlet in $54_H$ is ``D-even", $D(1,1,1)_{PS}=(1,1,1)_{PS}$, in contrast to the singlet in $210_H$ which is ``D-odd", $D(1,1,1)_{PS}=-(1,1,1)_{PS}$. So, if one breaks $SO(10)$ with the vacuum expectation value (vev) of $54_H$, D-parity is intact and one ends up getting $g_L=g_R$ at the energy scale associated with right-handed gauge boson ($W_R^\pm$), unlike the case of breaking initiated by the vev of $210_H$. As this discrete symmetry was used for the first time to decouple the Parity and $SU(2)_R$ breaking scales, it earned the name ``D-parity" \cite{Chang:1983fu}.}  ($SU(4)_C \times SU(2) _L \times SU(2)_R \times D$) while $16_H$ breaks the other $SO(10)$  to $SU(5)$ corresponding to a larger symmetry breaking which leaves an unwanted extra massless Goldstone boson in the $\left\lbrace ( 3,2,\frac{1}{6} ) +h.c. \right\rbrace$ representation of the SM gauge symmetry.\footnote{It will be interesting to study quantum salvation of such a model. We note, however, that generating large Majorana neutrino masses in this case would require introduction of new $SO(10)$ singlet fermions, which would make the model not so minimal.}

\begin{figure}[htb]
\begin{center}
\includegraphics[width=.8\textwidth]{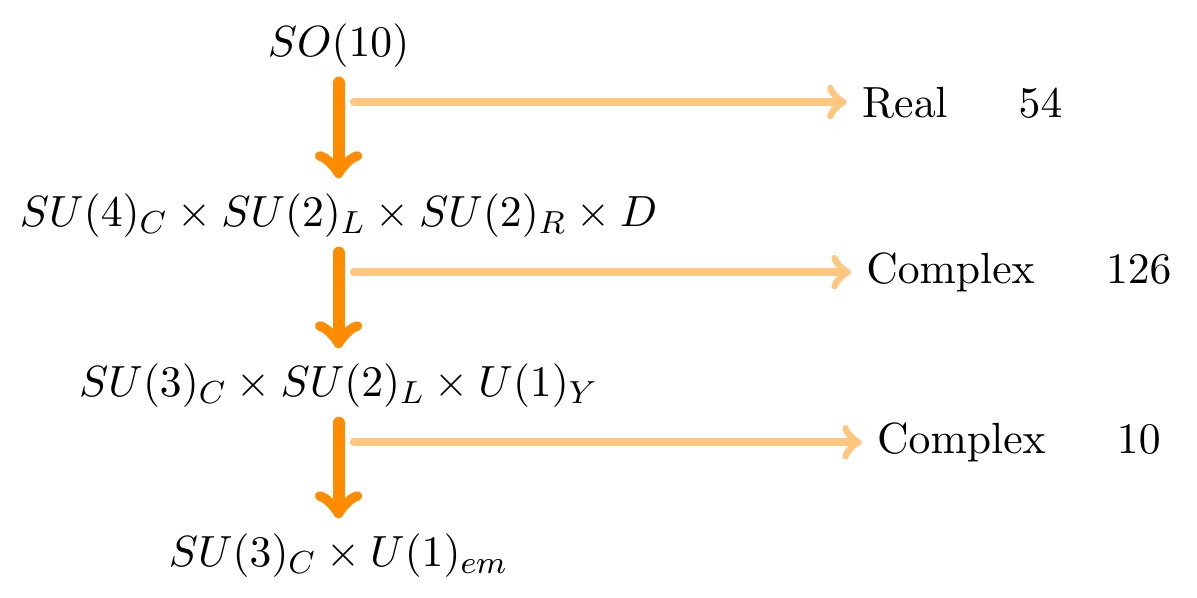}
\caption{$SO(10)$ symmetry breaking pattern with a Higgs sector consisting of $54_H+126_H + 10_H$.}
\label{Symmetry Breaking Pattern}
\end{center}
\end{figure}

In contrast, Higgs sector consisting a real $54_H$ and a complex $126_H$ along with a complex $10_H$ has all the properties that one needs to build a successful and predictive non-supersymmetric minimal $SO(10)$ GUT. In this case the symmetry breaking pattern is as shown in Fig. \ref{Symmetry Breaking Pattern}.

\noindent
In this breaking chain, 
\begin{enumerate}[label=(\emph{\roman*}),noitemsep, topsep=-10pt]
\item A real $54_H$ breaks the $SO(10)$ symmetry to the Pati-Salam symmetry with D-parity.
\item A complex $126_H$ breaks the Pati-Salam symmetry to the Standard Model.
\item A complex $10_H$ is used to perform the electroweak scale breaking.
\end{enumerate}
Even though a real $10_H$ is good enough to break the last stage of symmetry, a complex $10_H$ has been used here as the complex version is needed along with the complex $126_H$ to reproduce realistic fermion masses as argued in Ref. \cite{Bajc:2005zf}. The decomposition of the Higgs fields under the Pati-Salam (PS) group ($SU(4)_C \times SU(2)_L \times SU(2)_R$), the Standard Model group ($SU(3)_C \times SU(2)_L \times U(1)_Y$) and $SU(5) \times U(1)_X$ are given in the Table \ref{table:higgstable}.

\subsection{Associated Energy Scales}
In general, there are five possible energy scales associated with the left-right decomposition of any $SO(10)$ model, namely: (i) $M_U$  for $SO(10) \rightarrow SU(4)_C \times SU(2)_L \times SU(2)_R $, (ii) $M_C$ for $SU(4)_C \rightarrow SU(3)_C \times U(1)_{B-L} $, (iii) $M_R$ for $SU(2)_R \rightarrow U(1)_R$, (iv) $M'_R$ for $U(1)_R \times U(1)_{B-L}\rightarrow U(1)_Y$ and (v) $M_W$ for $SU(2)_L \times U(1)_Y \rightarrow U(1)_{em}$ \cite{Babu:1984mz}. In this minimal model, all three scales besides $ M_U$ and $M_W$ merge together and we will call the energy scale  $ M_C = M_R = M'_R \equiv M_I$, the intermediate scale. The presence of only one intermediate scale $(M_I)$ in between unified scale $(M_U)$ and the eletroweak scale $(M_W)$ makes the model highly constrained and predictive.

\textit{Extended Survival Hypothesis}: To ascertain the energy scales of all the Higgs multiplets, we evoke the philosophy known as ``Extended Survival Hypothesis". This is an extension of Geogri's ``Survival Hypothesis" for fermions \cite{Georgi:1979md} which states ``the representation that is invariant of the gauge group do acquire super-large mass". It was later extended to include scalar particles by the hypothesis: ``Higgs acquire the maximum mass compatible with the symmetry breaking."\cite{del Aguila:1980at, Mohapatra:1982aq}. Here we will employ a more relaxed version of the hypothesis by stating that ``Higgs multiplets remain at the maximum energy scale compatible with the symmetry breaking and phenomenological constraints." This is essentially a hypothesis of minimal fine-tuning.

According to this hypothesis, all the components of the $54_H$ remain at the unification scale, $M_U$. The $126_H$ decomposes as  $\Sigma_1 (6,1,1)_{PS} \oplus \Sigma_2 (10,3,1)_{PS} \oplus  \Sigma_3(\overline{10},1,3)_{PS} \oplus  \Sigma_4 (15,2,2)_{PS}$ under PS symmetry. As the PS symmetry is broken at the intermediate scale $M_I$ by the vacuum expectation value of $\left\langle\Sigma_3(\overline{10},1,3)_{PS}\right\rangle$, all the components of $\Sigma_3(10,1,3)_{PS}$ must remain at $M_I$. Due to the $D$-parity, the $\Sigma_2(10,3,1)_{PS}$ multiplet also remains at $M_I$. To reproduce all the realistic fermion masses and mixings, the $\Sigma_4(15,2,2)_{PS}$ needs to stay at the intermediate scale $M_I$. This is due to the fact that if $\Sigma_4(15,2,2)_{PS}$ lives at the unification scale, the induced electroweak vev of $\Sigma_4(15,2,2)_{PS}$ will get suppressed by the square of the ratio of intermediate scale and unification scale. In general such a small induced vev fails to correct the mass relations generated by only one complex $10_H$ Higgs \cite{Babu:1992ia}. These bad mass relations include $ V_{CKM}=  \mathds{1} $  and $m_u: m_c: m_t=m_d:m_s:m_b = m_e:m_\mu:m_\tau$. These wrong relations can be appropriately modified if the induced vev's are in the right order. Induced vev of order $m_b$ is needed for $\Sigma_4(15,2,2)_{PS}$ and a suppression of $\left(\nicefrac{M_I}{M_U}\right)^2 \sim 10^{-4}$ will be insufficient. Hence the need for $\Sigma_4(15,2,2)_{PS}$ being at $M_I$. As $126_H$ and $54_H$ have only one non-trivial cross coupling, the $\Sigma_1(6,1,1)_{PS}$ multiplet does not have enough freedom to be at the unification energy scale $M_U$ and will be brought down to $M_I$. This is a consequence of explicit Higgs potential analysis. In short, the whole $126_H$ has to be at the intermediate scale, $M_I$.

The weak scale breaking in this model is achieved by a complex $10_H$. So, only one of the doublets of SM needs to be at the weak scale. The other doublet needs to stay at the intermediate scale for the same reason as $\Sigma_4(15,2,2)_{PS}$. The rest of the multiplets (the color triplets) from $10_H$ will acquire masses of the order of the unification scale $M_U$.
\renewcommand\arraystretch{1.1}
\begin{table}[ph]
\centering
\begin{center}
\begin{tabular}{|c|c|c|c|c|}
 \hline 
 $\bm{SO(10)}$ & $\bm{SU(4)_C \times SU(2)_L \times SU(2)_R}$ & $\bm{SU(3)_C \times SU(2)_L \times U(1)_Y}$ &$\bm{SU(5) \times U(1)_X}$ & $\bm{Scale}$ \\ 
 \hline 
 \multirow{4}{*}{$10$}  & \multirow{2}{*}{$H_T (6,1,1)$} & $T_1 (3,1,-\frac{1}{3})$ &$(5,-2) $& $M_U$ \\ 
\hhline{~~---}
  &  & $T_2 (\overline{3},1,+\frac{1}{3})$ &$(\overline{5},+2) $& $M_U$ \\ 
 \hhline{~----} 
  & \multirow{2}{*}{$H_D (1,2,2)$} & $H_1 (1,2,-\frac{1}{2})$ &$(5,-2) $& $M_W$ \\ 
\hhline{~~---}
  &  & $H_2 (1,2,+\frac{1}{2})$ &$(\overline{5},+2) $& $M_I$ \\ 
\hline 
  \multirow{11}{*}{54}& \multirow{3}{*}{$\zeta_1(1,3,3)$} & $\zeta_{11} (1,3-1)$ &$(15,-4)$ & $M_U$\\
  \hhline{~~---}
  &  & $\zeta_{12} (1,3, 0)$ & $(24,0) $ & $M_U$\\
  \hhline{~~---}
  &  & $\zeta_{13} (1,3,+1)$ & $(\overline{15},+4) $ & $M_U$\\
   \hhline{~----}
   & \multirow{4}{*}{$\zeta_2(6,2,2)$} & \textcolor{red}{$\zeta_{21} (3,2,-\frac{5}{6})$} & \textcolor{red}{$(24,0) $} & \textcolor{red}{\bm{$M_U$}}\\
  \hhline{~~---}
  &  & \textcolor{red}{$\zeta_{22} (3,2,+\frac{1}{6})$}& \textcolor{red}{$(15, -4)$} & \textcolor{red}{\bm{$M_U$}}\\
  \hhline{~~---}
  &  & \textcolor{red}{$\zeta_{23} (\overline{3},2,-\frac{1}{6})$} &\textcolor{red}{ $(15, -4)$}& \textcolor{red}{\bm{$M_U$}}\\
  \hhline{~~---}
  &  & \textcolor{red}{$\zeta_{24} (\overline{3},2,+\frac{5}{6})$} & \textcolor{red}{$(24,0) $}& \textcolor{red}{\bm{$M_U$}}\\
  \hhline{~----}
   & \multirow{3}{*}{$\zeta_3(20',1,1)$} & $\zeta_{31} (\overline{6},1,+\frac{2}{3})$ & $(\overline{15},+4) $ & $M_U$\\
  \hhline{~~---}
  &  & $\zeta_{32} (6,1,-\frac{2}{3})$ &$(15,-4) $ & $M_U$\\
  \hhline{~~---}
  &  & $\zeta_{33} (8,1,0)$ & $(24,0) $ & $M_U$\\
 \hhline{~----}
 & $\zeta_0(1,1,1)$& $\zeta_{00} (1,1,0)$ & $(24,0) $ & $M_U$\\
\hline
  \multirow{22}{*}{126} & \multirow{2}{*}{$\Sigma_1 (6,1,1)$}&$\Sigma_{11} (3,1,-\frac{1}{3})$ & $(45,-2) $  & $M_U$\\
  \hhline{~~---}
   &  &$\Sigma_{12} (\overline{3},1,+\frac{1}{3})$ & $(\overline{5},+2) $& $M_U$\\
  \hhline{~----}
   &\multirow{3}{*}{$\Sigma_2 (10,3,1)$}&$\Sigma_{21} (1,3,-1)$ & $(\overline{15},-6)$ & $M_I$\\
  \hhline{~~---}
   &  &$\Sigma_{22} (3,3,-\frac{1}{3})$ & $(45,-2) $ & $M_I$\\
  \hhline{~~---}
   &  &$\Sigma_{23} (6,3,+\frac{1}{3})$ & $(\overline{50},+2) $&$M_I$\\
  \hhline{~----}
   &\multirow{9}{*}{$\Sigma_3 (\overline{10},1,3)$}&\textcolor{red}{$\Sigma_{31} (1,1,0)$ }&\textcolor{red}{ $(1,+10) $} & \textcolor{red}{\bm{$M_I$}}\\
  \hhline{~~---}
   &  &\textcolor{red}{$\Sigma_{32} (1,1,+1)$}&\textcolor{red}{$(10,+6)$} & \textcolor{red}{\bm{$M_I$}}\\
  \hhline{~~---}
   &  &$\Sigma_{33} (1,1,+2)$ &$(\overline{50},+2) $ & $M_I$\\
  \hhline{~~---}
   &  &$\Sigma_{34} (\overline{3},1,+\frac{4}{3})$ & $(10,+6) $ & $M_I$\\
  \hhline{~~---}
   &  &$\Sigma_{35} (\overline{3},1,+\frac{1}{3})$ & $(\overline{50},+2) $ & $M_I$\\  
  \hhline{~~---}
   &  &\textcolor{red}{$\Sigma_{36} (\overline{3},1,-\frac{2}{3})$} &\textcolor{red}{$(45,-2)$}& \textcolor{red}{\bm{$M_I$}}\\
  \hhline{~~---}
   &  &$\Sigma_{37} (\overline{6},1,-\frac{4}{3})$ & $ (\overline{50},+2)$& $M_I$\\
  \hhline{~~---}
   &  &$\Sigma_{38} (\overline{6},1,-\frac{1}{3})$ &$(45,-2) $ & $M_I$\\
  \hhline{~~---}
   &  &$\Sigma_{39} (\overline{6},1,+\frac{2}{3})$ &$(\overline{15},-6) $& $M_I$\\
  \hhline{~----}
   &\multirow{8}{*}{$\Sigma_4 (15,2,2)$}&$\Sigma_{41} (1,2,-\frac{1}{2})$ & $(\overline{5},+2) $& $M_I$\\
  \hhline{~~---}
   &  &$\Sigma_{42} (1,2,+\frac{1}{2})$ & $(45,-2) $ & $M_I$\\
  \hhline{~~---}
   &  &$\Sigma_{43} (3,2,+\frac{7}{6})$ & $(\overline{50},+2) $ & $M_I$\\
  \hhline{~~---}
   &  &$\Sigma_{44} (3,2,+\frac{1}{6})$ &$ (10,+6)$& $M_I$\\
  \hhline{~~---}
   &  &$\Sigma_{45} (\overline{3},2,-\frac{1}{6})$ &$(\overline{15},-6)$& $M_I$\\
  \hhline{~~---}
   &  &$\Sigma_{46} (\overline{3},2,-\frac{7}{6})$ &$(45,-2)$& $M_I$\\
  \hhline{~~---}
   &  &$\Sigma_{47} (8,2,-\frac{1}{2})$ &$(\overline{50},+2)$& $M_I$\\
  \hhline{~~---}
   &  &$\Sigma_{48} (8,2,+\frac{1}{2})$ &$(45,-2)$& $M_I$\\
 \hline
 \end{tabular}  
 \caption{Decomposition of the scalar representations with respect to various $SO(10)$ subgroups.  The ``scale" indicates expectation based on extended survival hypothesis. The Higgs multiplets in red (or bold) are the massless Goldstone bosons which are absorbed by the corresponding gauge bosons.}
 \label{table:higgstable}
 \end{center}
\end{table}
\section{Running of gauge couplings using two loop Renormalization Group Equations}
\label{sec:Running of gauge couplings using two loop Renormalization Group Equations}
In general, the three gauge couplings of the SM do not coincide at a single point when extrapolated using the SM renormalization group equation to high energy. But if a specific GUT, like the one under study, requires some Higgs bosons and gauge bosons other than the SM ones at a scale below $M_U$ and the newly introduced bosons have substantial effects on the beta functions, then it might be possible to assign suitable masses to these bosons and achieve unification of couplings. After specifying the Higgs sector of the model and the symmetry breaking pattern, one needs to run the couplings of the gauge groups with the appropriate beta functions and determine the status of the unification of the model under study. 

The two-loop renormalization group equations (RGE) for the gauge couplings can be written as:
\begin{equation}
\label{2loopRGE}
\dfrac{d \alpha^{-1}_i(\mu)}{d \ln \mu}=-\dfrac{a_i}{2 \pi}-\sum\limits_{j}\dfrac{b_{ij}}{8 \pi^2 \alpha^{-1}_j(\mu)} 
\end{equation}
where $i,j$ indices refer to different subgroups of the unified gauge group at the energy scale $\mu$ and 
\begin{equation}
\alpha^{-1}_i=\dfrac{4 \pi}{g_i^2}.
\end{equation}
\noindent
The $\beta$-function upto two-loop order is given by \cite{Machacek:1983tz}
\begin{eqnarray}
\beta (g) = \mu \dfrac{d g}{d\mu}&=&-\dfrac{g^3}{(4 \pi)^2}\left\lbrace\dfrac{11}{3}C_2 (G)-\dfrac{4}{3}\kappa S_2(F)-\dfrac{1}{6}\eta S_2(S)\right\rbrace  \nonumber\\ 
&& -\dfrac{g^5}{(4 \pi)^4}\left\lbrace \dfrac{34}{3} \left[ C_2(G)\right] ^2-\kappa\left[4 C_2(F) +\dfrac{20}{3}C_2(G)\right]S_2(F)\right. \nonumber\\ 
&& \left.-\left[2C_2(S)+\dfrac{1}{3} C_2(G)\right]\eta S_2(S)\right\rbrace . 
\end{eqnarray}
\noindent
Here $S_2$ and $C_2$ denote the Dynkin indices of the representations with the appropriate multiplicity factors (one has to be careful about whether the representation is complex or real) and the quadratic Casimir of a given representation. $\kappa=1,\frac{1}{2}$ for Dirac and Weyl fermions and $\eta=1,2$ for real and complex scalar fields. $G$, $F$ and $S$ stand for gauge multiplets, fermions and scalars.

\noindent
From the $\beta$-function expression we get,
\begin{eqnarray}
a_i&=&-\dfrac{11}{3}C_2 (G_i)+\dfrac{4}{3}\kappa S_2(F_i)+\dfrac{1}{6}\eta S_2(S_i)\\
b_{ij}&=&-\dfrac{34}{3} \left[ C_2(G_i)\right] ^2 \delta_{ij}+\kappa\left[4 C_2(F_j) +\dfrac{20}{3}\delta_{ij} C_2(G_i)\right]S_2(F_i) \nonumber\\ 
&&+\eta \left[2C_2(S_j)+\dfrac{1}{3} \delta_{ij}  C_2(G_i)\right]S_2(S_i) .
\end{eqnarray}

\noindent
The one-loop and two-loop $\beta$-function coefficients for the Standard Model (valid for $M_W \leq \mu \leq M_I$), and the Pati-Salam symmetry group with D-parity (valid for $M_I \leq \mu \leq M_U$) are found to be
\renewcommand\arraystretch{1.2}
\begin{equation}
\label{eq:beta coefficients SM}
a_{SM}=\left(\begin{array}{r}
-7 \\
-\frac{19}{6} \\ 
\frac{41}{10} 
\end{array} \right);\hspace{1cm} b_{SM}=\left( \begin{array}{rcc}
-26 & \frac{9}{2}& \frac{11}{10} \\
12 & \frac{35}{6} & \frac{9}{10}\\
\frac{44}{5}& \frac{27}{10} & \frac{199}{50}
\end{array} \right) ;
\end{equation}

\begin{equation}
\label{eq:beta coefficients PS}
a_{PS}=\left(\begin{array}{c}
1 \\
\frac{26}{3} \\ 
\frac{26}{3} 
\end{array} \right);\hspace{1cm} b_{PS}=\left( \begin{array}{ccc}
\frac{1209}{2} & \frac{249}{2} & \frac{249}{2} \\
\frac{1245}{2} & \frac{779}{3} & 48 \\
\frac{1245}{2} & 48 & \frac{779}{3}
\end{array} \right) .
\end{equation}
\noindent
Here we have used the intermediate scale  scalar spectrum of Table \ref{table:higgstable}, along with the intermediate gauge symmetry $SU(4)_C \times SU(2)_L \times SU(2)_R \times D$.

The appropriate matching conditions for two-loop RGE, when a simple gauge group $\mathcal{G}$ spontaneously breaks down into subgroups $\mathcal{G}_i$'s, is given by \cite{Hall:1980kf}
\begin{equation}
\label{eq:Threshold Def}
\dfrac{1}{\alpha_i(\mu)}=\dfrac{1}{\alpha_G(\mu)}-\dfrac{\lambda_i(\mu)}{12 \pi} ,
\end{equation}
where
\begin{eqnarray}
\lambda_i(\mu)&=& \overbrace{\left( C_G-C_i \right)}^{\lambda_i^G} -21 \; \overbrace{Tr\left( t_{iV}^2 \ln \frac{M_V}{\mu}\right)}^{\lambda_i^V} \nonumber\\ 
&&  +\underbrace{Tr \left(t_{iS}^2 P_{GB} \ln \dfrac{M_S}{\mu} \right)}_{\lambda_i^S} + 8 \;\underbrace{Tr \left(t_{iF}^2  \ln \dfrac{M_F}{\mu} \right) }_{\lambda_i^F} .
\end{eqnarray}
\noindent
Here $V$, $F$ and $S$ denote respectively vector, fermion and scalar particles that are integrated out at the matching scale $\mu$; $C_G$ and $C_i$ denote the quadratic Casimir invariants of the groups $\mathcal{G}$ and  $\mathcal{G}_i$; $t_{i \left\lbrace V,F,S\right\rbrace }$'s are the generators of the lower symmetry $\mathcal{G}_i$ for the representations in which the heavy \{Gauge bosons, fermions, scalar bosons\} appear; $P_{GB}$ is a projection operator which projects out all the Goldstone bosons.

Equipped with all these RGE's and matching conditions given above, it becomes straightforward to determine the intermediate scale $M_I$ and the unification scale $M_U$ for our model. Let us first find out these scales completely ignoring the threshold corrections stemming from the gauge bosons and unknown masses of the Higgs particles. In this scenario, we assume that all the Higgs and gauge boson masses are degenerate with masses equal to either $M_I$ or $M_U$ as dictated by the extended survival hypothesis. For the group $U(1)_Y$ the appropriate matching condition is given by
\begin{equation}
\dfrac{1}{\alpha_{1Y}(\mu)}=\dfrac{3}{5}\left(\dfrac{1}{\alpha_{2R}(\mu)}-\dfrac{C_{2R}}{12 \pi}\right)+\dfrac{2}{5}\left(\dfrac{1}{\alpha_{4C}(\mu)}-\dfrac{C_{4C}}{12 \pi}\right).
\end{equation}
 
\begin{figure}[!htb]
\centering
\includegraphics[scale=.3]{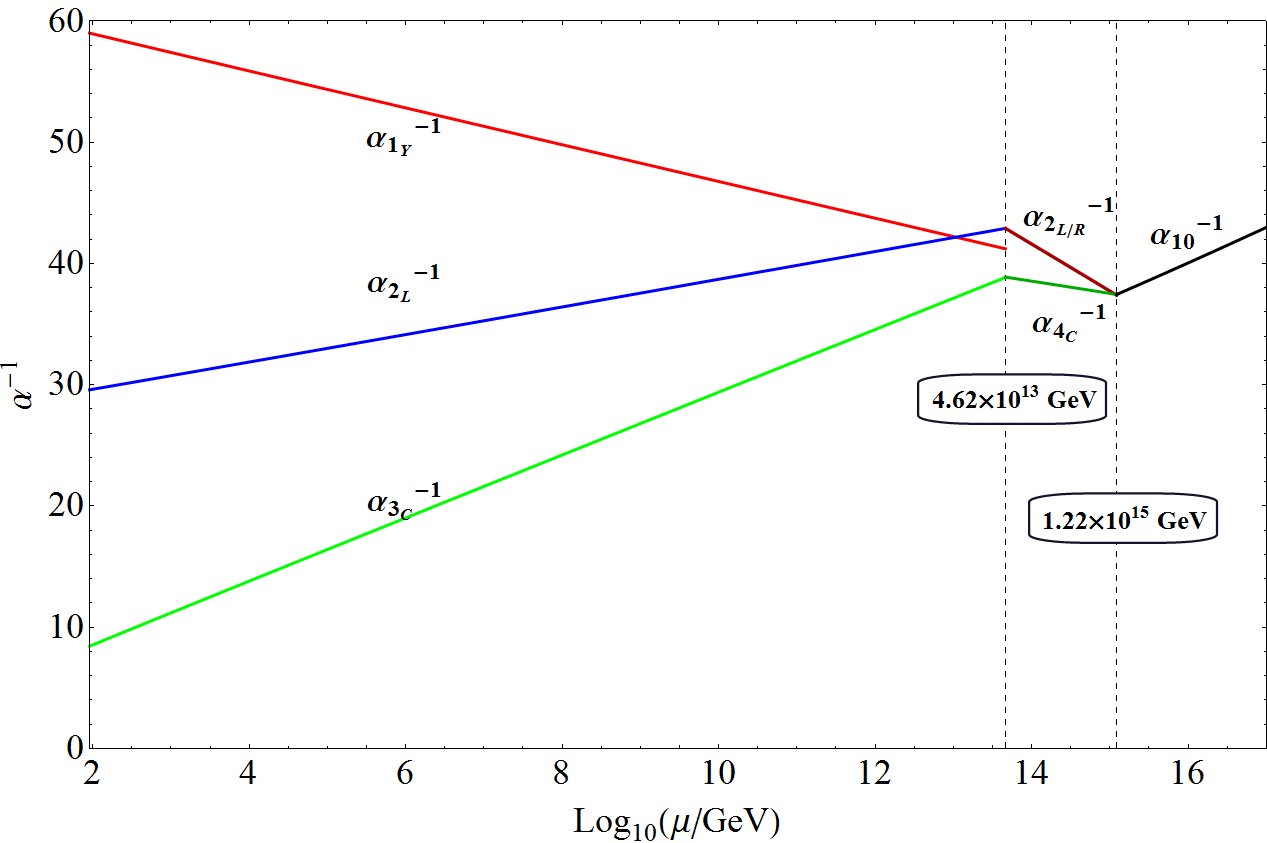}
\caption{Running of gauge couplings without threshold corrections using two-loop RGE. Pati-Salam symmetry with D parity is assumed as the intermediate scale. The dotted vertical lines correspond to the intermediate scale and the unification scale. }
\label{fig:Running of Gauge Coupling without Threshold Correction}
\end{figure} 

As input at $\mu=M_Z$, we use 
\begin{align}
\label{eq:lowdata}
\alpha_{1Y}^{-1}&= \nicefrac{3}{5} \;  \alpha_{em}^{-1} (1-\sin^2\theta); \vspace{.5cm}&\vspace{.5cm} \alpha_{2L}^{-1}&=\alpha_{em}^{-1} \sin^2\theta ;\nonumber\\
M_Z &= 91.1876 \; \text{GeV}; \vspace{.5cm}&\vspace{.5cm} \alpha_{em}^{-1}&=127.940;\nonumber\\
\sin^2\theta &=0.23126; \vspace{.5cm}&\vspace{.5cm}  \alpha_{3c}^{-1}&=0.1185 .
\end{align}

By solving the two-loop RGE numerically, one obtains the intermediate scale to be $4.62 \times 10^{13}\; \text{GeV}$ and the unification scale to be $1.22 \times 10^{15}\; \text{GeV}$. Such a unification scale is obviously ruled out by the current bound on proton lifetime, $ \tau_p(p \rightarrow e^+ \pi^0) > 1.29 \times 10^{34}\; \text{yrs}$ which requires $M_U \gtrsim (5 \sim 7) \times 10^{15}\;\text{GeV}$ \cite{Nishino:2012ipa}. It is this feature that has made the model less studied. But as we show in the next section, threshold corrections arising from the scalar sector at $M_I$  and $M_U$ can nicely rectify the situation and make the model consistent and still testable.

\section{Threshold corrections}
\label{sec:Threshold Correction}
A fundamental limitation of all GUTs is the lack of underlying physics required to predict the masses of Higgs bosons and thus the threshold corrections associated with them. To improve the situation somewhat, one can always go through the tedious process of writing down the whole Lagrangian for the $SO(10)$ model and then derive the scalar particle mass spectrum in terms of the couplings and vev's of the $SO(10)$ Lagrangian. In such a scenario, the couplings and vev's are only constrained by the phenomenology and perturbitivity arguments.

Even though the scale of the physics, masses of the gauge bosons and the Higgs bosons should be around the same order of energy, nothing dictates that they should all be exactly the same. On a stronger note, one can say that even though some of them can be degenerate, it is unlikely that all of them are. And this distribution of masses and vev's generates the threshold corrections. This  becomes a very important factor for Higgs bosons belonging to a large representation such as $126_H$. The predictions of the model derived while ignoring the threshold corrections becomes unreliable. 

Here we proceed to take account of these corrections and study the resulting modification of the values of $M_I$ and $M_U$. At first we include the threshold corrections in a very generic manner. At this stage, we assume that all the Higgs masses are independent and they are selected in a completely random manner. The only constraint that we put on the Higgs masses is due to the extended survival hypothesis.  We kept the ratio, $R$ of Higgs boson mass and the corresponding gauge boson masses to be between $R=\left\lbrace\nicefrac{1}{10},\nicefrac{1}{20},\nicefrac{1}{33}\right\rbrace$ and $R=2$. While ignoring any relations between the scalar masses, this analysis gives us the maximum threshold correction to the energy scales in terms of the spread of the distribution of scalar masses. 

\subsection*{Defining the Pati-Salam Scale $M_{PS}$}
Low energy experimental data fixes the initial points for the running of the gauge couplings while demand of unification puts a couple of constraints on the evolution. These determine two unique scales of the model. Let us consider the running of the three gauge couplings upto a scale $\Lambda_i$ which is higher than the heaviest component of $126_H$ but lighter than any component of $54_H$. For such a situation we have:
\begin{eqnarray}\label{eg: Defining PS Scale}
\alpha^{-1}_{3C}(M_Z) &=& \alpha^{-1}_{4C}(\Lambda_i)+\dfrac{a_{3C}}{2 \pi} \ln \left[ \dfrac{\Lambda_i}{M_Z}\right]-\dfrac{\lambda_{3C}^i}{12 \pi};
  \nonumber\\
\alpha^{-1}_{2L}(M_Z) &=& \alpha^{-1}_{2L}(\Lambda_i)+\dfrac{a_{2L}}{2 \pi} \ln\left[ \dfrac{\Lambda_i}{M_Z}\right] -\dfrac{\lambda_{2L}^i}{12 \pi};
 \nonumber\\
\alpha^{-1}_{1Y}(M_Z) &=& \frac{3}{5}\alpha^{-1}_{2R}(\Lambda_i)+\frac{2}{5}\alpha^{-1}_{4C}(\Lambda_i)+\dfrac{a_{1y}}{2 \pi} \ln \left[\dfrac{\Lambda_i}{M_Z} \right]-\dfrac{\lambda_{1Y}^i}{12 \pi} .
\end{eqnarray}
With the notation $\eta_j^a=ln\dfrac{M_j}{M_a}$; $j$ being any Higgs multiplet and $a=i,u$ being intermediate scale ($M_I$) or unification scale ($M_U$) respectively:
\begin{eqnarray}\label{eq:Threshold at Mi S+G}
\lambda_{3C}^{iS} &=& \eta_{\Sigma_{11}}^i+\eta_{\Sigma_{12}}^i+3 \eta_{\Sigma_{22}}^i + 15\eta_{\Sigma_{23}}^i + \eta_{\Sigma_{34}}^i+ \eta_{\Sigma_{35}}^i  + 5\eta_{\Sigma_{37}}^i +5 \eta_{\Sigma_{38}}^i + 5 \eta_{\Sigma_{39}}^i   \nonumber\\ &&  +2 \eta_{\Sigma_{43}}^i + 2\eta_{\Sigma_{44}}^i+ 2 \eta_{\Sigma_{45}}^i + 2 \eta_{\Sigma_{46}}^i + 12 \eta_{\Sigma_{47}}^i + 12\eta_{\Sigma_{48}}^i; \nonumber \\ [0.5em]
\lambda_{2L}^{iS}&=& \eta_{H_2}^i + 4\eta_{\Sigma_{21}}^i+12 \eta_{\Sigma_{22}}^i +24 \eta_{\Sigma_{23}}^i+\eta_{\Sigma_{41}}^i +\eta_{\Sigma_{42}}^i +3 \eta_{\Sigma_{43}}^i + 3\eta_{\Sigma_{44}}^i + 3\eta_{\Sigma_{45}}^i \nonumber \\&& + 3\eta_{\Sigma_{46}}^i +8\eta_{\Sigma_{47}}^i +8\eta_{\Sigma_{48}}^i; \nonumber \\ [0.5em]
\lambda_{1Y}^{iS} &=&\frac{1}{5} \left( 3 \eta_{H_2}^i +2\eta_{\Sigma_{11}}^i+2\eta_{\Sigma_{12}}^i+ 18\eta_{\Sigma_{21}}^i +6\eta_{\Sigma_{22}}^i +12\eta_{\Sigma_{23}}^i +24\eta_{\Sigma_{33}}^i +32\eta_{\Sigma_{34}}^i+2\eta_{\Sigma_{35}}^i   \right. \nonumber \\  && \hspace{.5cm} \left. +64\eta_{\Sigma_{37}}^i +4\eta_{\Sigma_{38}}^i+16\eta_{\Sigma_{39}}^i +3\eta_{\Sigma_{41}}^i + 3\eta_{\Sigma_{42}}^i + 49\eta_{\Sigma_{43}}^i+  \eta_{\Sigma_{44}}^i  + \eta_{\Sigma_{45}}^i + 49\eta_{\Sigma_{46}}^i \right. \nonumber \\ && \hspace{.5cm} \left.+24\eta_{\Sigma_{47}}^i +24\eta_{\Sigma_{48}}^i \right); \nonumber \\
\lambda_{3C}^{iV} &=&  \eta_{PSV}^i; \hspace{.25cm}
\lambda_{2L}^{iV} = 0; \hspace{.25cm}
\lambda_{1Y}^{iV} = \dfrac{8}{5} \eta_{PSV}^i+\dfrac{6}{5} \eta_{W_R}^i; \hspace{.25cm}
\lambda_{3C}^{iG} = 1; \hspace{.25cm}
\lambda_{2L}^{iG} = 0; \hspace{.25cm}
\lambda_{1Y}^{iG} = \dfrac{14}{5}. 
\end{eqnarray}
\noindent 
Here $PSV$ is the Pati-Salam gauge boson$ (\overline{3},1,-\frac{2}{3})$ and $W_R$ is the right-handed $W_R^\pm (1,1,-1)$. The coefficients of the $\eta$'s are the Dynkin indices of the representations of the respective gauge group together with the multiplicity factors. For the case of hypercharge GUT-compatible normalization has been used.
\noindent
Along with Eq. (\ref{eq:Threshold at Mi S+G}) we find the following equation:
\begin{equation}\label{eq:Mps_def}
\begin{split}
\left(5 \alpha^{-1}_{1Y}-3 \alpha^{-1}_{2L} - 2 \alpha^{-1}_{3C}  \right) (M_Z) & =  \frac{1}{2 \pi} \left\lbrace - 2 + \ln\dfrac{ M_{PS}^{44}}{M_Z^{44} }\right\rbrace
\end{split}
\end{equation}
where
\begin{equation}\label{eq:Mps_def2}
M_{PS}=\left(\dfrac{ M_{PSV}^{21} M_{W_R}^{21} M_{\Sigma_{22}}^6 M_{\Sigma_{23}}^{15} M_{\Sigma_{38}} M_{\Sigma_{44}}^2 M_{\Sigma_{45}}^2 M_{\Sigma_{47}}^4 M_{\Sigma_{48}}^4}{ M_{\Sigma_{21}} M_{\Sigma_{33}}^4 M_{\Sigma_{34}}^5 M_{\Sigma_{37}}^9 M_{\Sigma_{39}} M_{\Sigma_{43}}^6 M_{\Sigma_{46}}^6}\right)^{\nicefrac{1}{44}} .
\end{equation}
\noindent
Here $M_{PSV}$ is the mass of the Pati-Salam gauge boson$ (\overline{3},1,-\frac{2}{3})$ and $M_{W_R}$ is the mass of right-handed $W_R^\pm (1,1,-1)$. 
Eq. (\ref{eq:Mps_def}) completely determines the ``Pati-Salam Scale ($M_{PS}$)" in terms of low energy experimental data, which will be the intermediate scale unless otherwise mentioned.

Using one-loop RGE, we find the intermediate scale (Pati-Salam scale) to be:
\begin{equation}
M_{PS}^{1-loop} = M_Z \; e^{\frac{C_{ps}}{44}} = 5.33 \times 10^{13} \; GeV  \nonumber
\end{equation}
where 
\begin{equation}
C_{ps} = 2 \pi \left( 5 \alpha^{-1}_{1Y}(M_Z)-3 \alpha^{-1}_{2L}(M_Z) - 2 \alpha^{-1}_{3C}(M_Z) \right)+2  \nonumber
\end{equation}
and we have used the data given in Eq. (\ref{eq:lowdata}).
\noindent
To reduce the error coming from the fact that this definition does not use two-loop RGE running, we can run the SM gauge couplings at two-loop level upto the energy $\approx 10^{12}$ GeV. In that case, we find that $M_{PS} = 4.67 \times 10^{13} $~GeV which is very close to the value obtained by using two-loop RGE running upto unification scale.

Analytically it is tricky to define the Pati-Salam/intermediate scale at the two-loop level. Nevertheless, as the scales should not depend on the threshold corrections, one can evolve the couplings at two-loop level assuming all the scalar and gauge bosons to be degenerate at the respective scales. In that case, the unification constraint and low energy data fix the scale to be $M_{PS} = 4.62 \times 10^{13} $~GeV. This indicates an important fact that, if we consider two-loop RGE upto an energy scale $\approx 10^{12}\; \text{GeV}$ and then analyze the rest of the evolution (upto the unification scale) at one-loop level, the error introduced should not change the result drastically. The two-loop effects cannot accumulate a large amount of corrections in the process of running by only three orders of magnitude in energy scale from $(10^{12} - 10^{15})\;\text{GeV}$.
\subsection*{Defining the unification scale $M_U$}
After defining the PS Scale, we can forget about the threshold corrections at the intermediate scale  and use the new-found PS scale for any calculation needed for determining the unification scale. So, starting from the Pati-Salam scale we can write a new set of RGE for the couplings at an energy scale $\Lambda_u$ higher than the energy scale corresponding to all the scalar particle masses as:
\begin{eqnarray}\label{eg: Defining GUT Scale}
\alpha^{-1}_{4C}(M_{PS}) &=& \alpha^{-1}_U(\Lambda_u)+\dfrac{a_{4C}}{2 \pi} \ln \left[ \dfrac{\Lambda_u}{M_{PS}}\right] -\dfrac{\lambda_{4C}^u}{12 \pi};
 \nonumber\\
\alpha^{-1}_{2L}(M_{PS}) &=& \alpha^{-1}_U(\Lambda_u)+\dfrac{a_{2L}}{2 \pi} \ln\left[ \dfrac{\Lambda_u}{M_{PS}}\right] -\dfrac{\lambda_{2L}^u}{12 \pi};
 \nonumber \\
\alpha^{-1}_{2R}(M_{PS}) &=& \alpha^{-1}_U(\Lambda_u)+\dfrac{a_{2R}}{2 \pi} \ln\left[ \dfrac{\Lambda_u}{M_{PS}}\right] -\dfrac{\lambda_{2R}^u}{12 \pi}.
\end{eqnarray}

Here
\begin{equation}
\label{eq:Threshold at Mu S+G}
\lambda_{4C}^{uS} = 2 \eta_{H_T}+8\eta_{\zeta_3}; \hspace{.5cm} 
\lambda_{2L}^{uS} = 6 \eta_{\zeta_1}; \hspace{.5cm} 
\lambda_{2R}^{uS} = 6 \eta_{\zeta_1}; \nonumber
\end{equation}
\begin{equation}
\lambda_{4C}^{uV} = 4 \eta_{uV}; \hspace{.25cm} 
\lambda_{4C}^{uG} = 4; \hspace{.25cm} 
\lambda_{2L}^{uV} = 6 \eta_{uV}; \hspace{.25cm} 
\lambda_{2L}^{uG} = 6; \hspace{.25cm} 
\lambda_{2R}^{uV} = 6 4 \eta_{uV}; \hspace{.25cm}
\lambda_{2R}^{uG} = 6
\end{equation}
\noindent
where the notation $uV$ corresponds to the leptoquark gauge boson at the unification scale ($M_U$). Using these equations, we find:
\begin{equation}\label{eq:Unification scale}
\begin{split}
 \alpha^{-1}_{4C}(M_{PS}) - \alpha^{-1}_{2L}(M_{PS})&=\dfrac{1}{6 \pi} \left\lbrace 1- \ln \frac{M_{U}^{23}}{M_{PS}^{23}} \right\rbrace
\end{split}
\end{equation}
\noindent
where
\begin{equation}\label{eq:Unification scale2}
M_U=\left(\frac{M_{uV}^{21}M_{H_T} M_{\zeta_3}^4}{ M_{\zeta_1}^3} \right)^{\frac{1}{23}}.
\end{equation}

Just like the Pati-Salam scale, we can define a unification scale ($M_U$) completely fixed by the Pati-Salam scale:
\begin{equation}
M_U^{1-loop}=M_{PS} \; e^{\frac{C_u}{23}}= 2.4 \times 10^{15}\; GeV
\end{equation}
\noindent
where
\begin{equation}
C_u= 1- 6 \pi \left( \alpha_{4C}(M_{PS})-\alpha_{2L}(M_{PS})\right) . \nonumber
\end{equation}

Similarly to the Pati-Salam scale, if one runs SM gauge couplings at two-loop level to an energy scale of $10^{12}$~GeV, the unification scale becomes $M_U=1.36\times 10^{15}$~GeV. And at two-loop level we find $M_U^{2-loop}=1.22\times 10^{15}$~GeV. Again, this small discrepancy in $M_U$ is due to the fact that the latter one is a two-loop gauge coupling evolution upto the unification scale while the previous one is two-loop level upto an energy level of $10^{12}$~GeV.

\subsection*{Threshold corrections at the unification scale}
After defining the scales and finding out the scales of all the Higgs bosons (given in Table \ref{table:higgstable}), it is straightforward to calculate analytically  the threshold corrections for the one-loop running of the gauge couplings \cite{Mohapatra:1992dx}. Numerically it is possible to improve the process, by using two-loop RGE (Eq. (\ref{2loopRGE})) and one-loop threshold correction formulas given in  Eqs. (\ref{eq:Threshold Def}), (\ref{eq:Threshold at Mi S+G}), (\ref{eq:Threshold at Mu S+G}). 

It is obvious that the masses of the gauge bosons will depend on the randomness adopted for the heavy Higgs masses due to the unification constraints given by Eqs. (\ref{eq:Mps_def}), (\ref{eq:Unification scale}). Guided by the extended survival hypothesis, we decided to allow the ratio of the mass of each Higgs boson to the corresponding gauge boson mass to be between $R=\left\lbrace\nicefrac{1}{10},\nicefrac{1}{20},\nicefrac{1}{33}\right\rbrace$  and $2$, with $R = \dfrac{M_{\text{Higgs boson}}}{\text{Corresponding gauge boson mass}}$. We study three cases where $R^{\text{-}1}=10,20$ and $33$. The upper limit of $2$ comes from the fact that we do not want to risk the perturbitivity of the model.

For a random sample of Higgs masses lying within the pre-selected range, one finds out the one-loop threshold corrections. Then using two-loop RGE for running of the gauge couplings and the unification constraints one determines the masses of the gauge bosons. Using the newly obtained gauge boson masses as updated scales, one repeats the process. After a few iterations, one ends up with a two-loop RGE of gauge couplings with one-loop threshold corrections with intermediate scale and unification scale at the corresponding gauge boson masses. Fig. \ref{fig:Running of Gauge Coupling without Threshold Correction} is a sample of such running of gauge couplings using two loop RGE and one loop threshold corrections.

\begin{figure}[!htb]
\centering
\includegraphics[width=.8\textwidth]{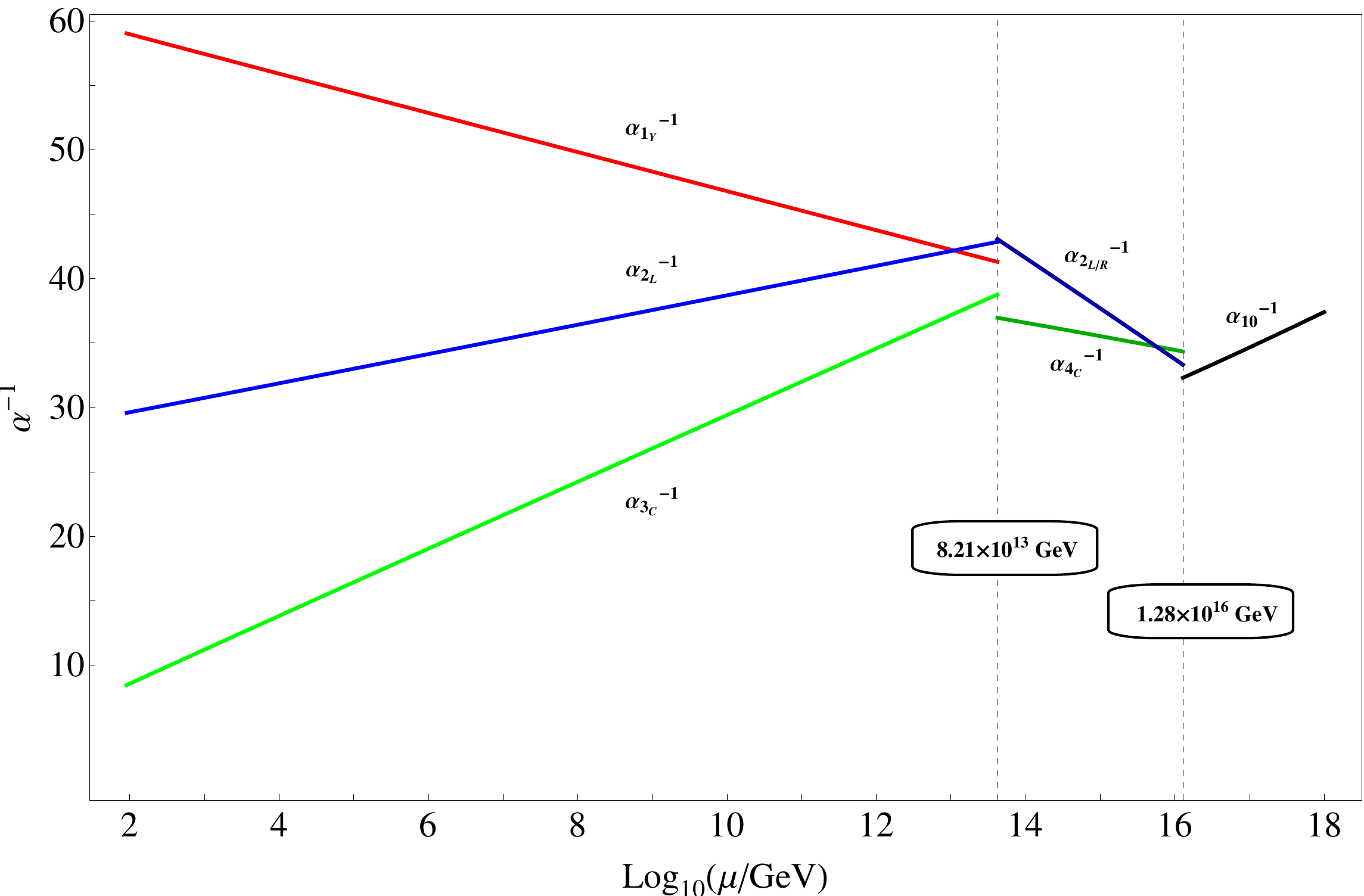}
\caption{Running of gauge couplings with one-loop threshold corrections using two-loop RGE. This sample point corresponds to a case where some of the Higgs masses are taken to be two times the corresponding gauge boson mass determined without the threshold corrections and the others are one tenth of the scale. Special attention was given to the color triplet masses, so that they are heavier than $10^{13}\; \text{GeV}$. In this extreme scenario, the mass of the leptoquark gauge boson (the one responsible of proton decay) is maximized. Then the scales were updated with an iteration process so that the scales correspond to the masses of the respective gauge bosons.}
\label{fig:Running of Gauge Coupling without Threshold Correction}
\end{figure}

To find out the pattern of the Higgs boson masses allowed by the current experimental bound, one needs to find out the proton lifetime in terms of the masses of the leptoquark gauge boson and the unified gauge coupling. 

\section{Proton Lifetime}
\label{sec:Proton Lifetime}
\textbf{Gauge mediated proton decay}: In non-supersymmetric GUTs, the primary mode of proton decay is $p \rightarrow e^+ \pi^0$. This gauge induced $d=6$ operator predicts a proton lifetime of the order of $\tau_p \approx \nicefrac{M_{(X,Y)}^4}{\left(g^4 m_p^5\right)}$, where $M_{(X,Y)}$ is the mass of the leptoquark (known as $X,Y$ gauge boson), $g^2 \approx \dfrac{4 \pi}{35}$ is the coupling at unification energy and $m_p $ is the mass of the proton.

We have calculated the rates for proton decay using a more detailed version of the lifetime formula which includes the relevant hadronic matrix elements of operators and also renormalization effect of the operator in going from GUT energy scale to $\mu = 1\; \text{GeV}$. The proton lifetime formula in $SO(10)$ models is \cite{Nath:2006ut, Bertolini:2013vta}:
\begin{equation}
\label{eq:proton lifetime formula}
\tau_P = \left[ \dfrac{\pi}{4}  \dfrac{m_p \alpha_U^2}{f_\pi^2} \left|\alpha_H \right|^2 R_L^2 (1+F+D)^2 \left(A_{SR}^2\left(\dfrac{1}{M^2_{(X,Y)}}+\dfrac{1}{M^2_{(X',Y')}}\right)^2+\dfrac{4 A_{SL}^2}{M^4_{(X,Y)}}\right)\right]^{-1} 
\end{equation}
where $R_L=1.36$ is the two-loop long-range running effect on the effective proton decay operator \cite{Nihei:1994tx}, $\alpha_H \simeq -0. 01 GeV^3$ \cite{Aoki:2006ib} denotes the relevant hadronic matrix element defined by $\alpha_H u^p_L(\vec{k}) \equiv \epsilon_{\alpha \beta \gamma}  \bra{0}d^\alpha_R u^\beta_Ru^\gamma_L \ket{p(\vec{k})}$, $D=0.8$ and $F=0.47$ are chiral Lagrangian parameters \cite{Aoki:2006ib}, $f_\pi =130.7\;\text{MeV}$ is the pion decay constant and $g_G$ is the gauge coupling constant at the unification scale. $M_{(X,Y)}$ and $M_{(X',Y')}$ are the masses of the corresponding gauge bosons,  $(X,Y)(3,2,+\nicefrac{5}{6})$ and $(X',Y')(3,2,-\nicefrac{1}{6})$. $A_{SL(R)}$ is the short-range left-handed (right-handed) renormalization factors of the proton decay operator corresponding to the running from the scale $\mu= M_U$ to $M_Z$, passing through the intermediate scale $M_I$ and given by
\begin{equation}
A_{SL(R)} = \prod_{i=1}^n \prod_{sc}^{M_Z \leq m_{sc} < M_U} \left[ \dfrac{\alpha_i(m_{sc+1})}{\alpha_i(m_{sc})}\right]^{\dfrac{\gamma_{L(R)(sc)i}}{a_i(m_{sc+1}-m_{sc})}}\nonumber
\end{equation}
where 
\begin{equation}
\gamma_{L (M_Z)} = \left\lbrace \dfrac{23}{20},\dfrac{9}{4},2\right\rbrace ; \hspace{.5cm}
\gamma_{R (M_Z)} = \left\lbrace \dfrac{11}{20},\dfrac{9}{4},2 \right\rbrace ; \hspace{.5cm}
\gamma_{L/R (M_{PS})} = \left\lbrace \dfrac{15}{4},\dfrac{9}{4},\dfrac{9}{4}\right\rbrace. \nonumber
\end{equation}
And $a_i$'s are the one-loop beta-function coefficients given in Eqs (\ref{eq:beta coefficients SM}) -(\ref{eq:beta coefficients PS}) and the relevant scales ($sc$) are $M_U, M_{PS}, M_Z$ ($sc=1,2,3$). 

As the $SO(10)$ model under scrutiny has a realistic and predictive Yukawa sector, the fermion masses and mixings can be determined by some fitting algorithm \cite{Joshipura:2011nn, Altarelli:2013aqa}. After one gets the explicit numerical values for the fermion masses and mixings, it is trivial to determine the branching ratios of various proton decay channels. The issue will be discussed in details in Sec. \ref{sec: Proton branching ratio} after we analyze the Yukawa sector of the model. 

\textbf{Higgs boson mediated proton decay}: The Higgs boson induced $d=6$ proton decay operator has the potential to play a vital role in the proton lifetime determination if certain scalar color triplets $T(3,1,-\nicefrac{1}{3})$ become light enough. As mentioned earlier, not all the dangerous Higgs color triplets are at the unification scale. A couple of them slide down to the intermediate scale due to the fact that the 54 and 126 plet of $SO(10)$ have only one non-trivial cross coupling which is fine-tuned to keep the $(15,2,2)_{PS}$ at the intermediate scale. 

In general, Higgs boson induced $d=6$ operators are suppressed by the first generation Yukawa couplings.($ \tau_p \approx \nicefrac{m_T^4}{\left| Y_u Y_d\right|^2 m_p^5}$) \cite{Nath:2006ut}. As, in the SM, $\left( Y_u Y_d \right) \approx 10^{-10}$, in all cases we took a conservative lower limit and kept all the Higgs color triplet  $T(3,1,-\nicefrac{1}{3})$ mass, $m_T > 10^{13}$ GeV, so that they do not contribute significantly to proton decay.

\textbf{Proton lifetime and threshold corrections:} Proton lifetime is very sensitive to the mass of the leptoquark gauge boson which in turn depends on the randomness adopted for the heavy Higgs masses. After we find out the masses of the gauge bosons and the unified gauge coupling using one-loop threshold corrections and two-loop RGE running of the gauge couplings, we can use Eq. (\ref{eq:proton lifetime formula}) to determine proton lifetime.

From our numerical analysis, we find that one can maximize proton lifetime by maximizing the masses of the following Higgs bosons: $\Sigma_{21},\Sigma_{22},\Sigma_{23},\Sigma_{33},\Sigma_{34},\Sigma_{41},\Sigma_{42},\Sigma_{43}$, $\Sigma_{44}$, $\Sigma_{45}$, $\Sigma_{46}$, $H_2$, $\zeta_{11},\zeta_{12},\zeta_{13} $ and by minimizing: $T_1, T_2,\Sigma_{11},\Sigma_{12},\Sigma_{35},\Sigma_{37}, \Sigma_{38}, \Sigma_{39},\Sigma_{47},\Sigma_{48}$, $\zeta_{31}$, $\zeta_{32}$, $\zeta_{33}$. In such a maximal/minimal arrangement, the proton lifetime can go as large as:
\begin{equation}
\begin{split}
\tau_{max} & = 1.45 \times 10^{35} \; \text{yrs}; \hspace{1cm} R^{\text{-}1}=10, \\
\tau_{max} & = 9.85 \times 10^{35} \; \text{yrs}; \hspace{1cm} R^{\text{-}1}=20, \\
\tau_{max} & = 3.91 \times 10^{36} \; \text{yrs}; \hspace{1cm} R^{\text{-}1}=33. 
\end{split}
\end{equation}
\begin{figure}
\centering
\begin{subfigure}{.5\textwidth}
 \centering
 \includegraphics[width=.95\linewidth]{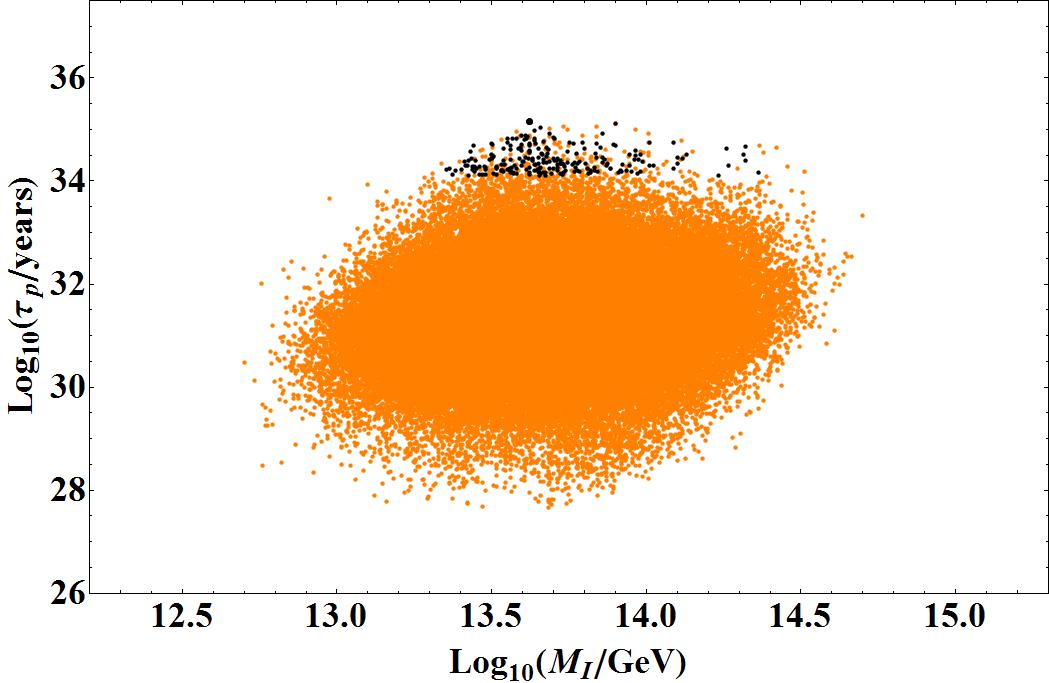}
  \caption{$R^{-1}=10$}
  \label{fig:proton lifetime vs Mps Rt=10}
\end{subfigure}%
\hfill
\begin{subfigure}{.5\textwidth}
  \centering
  \includegraphics[width=0.95\linewidth]{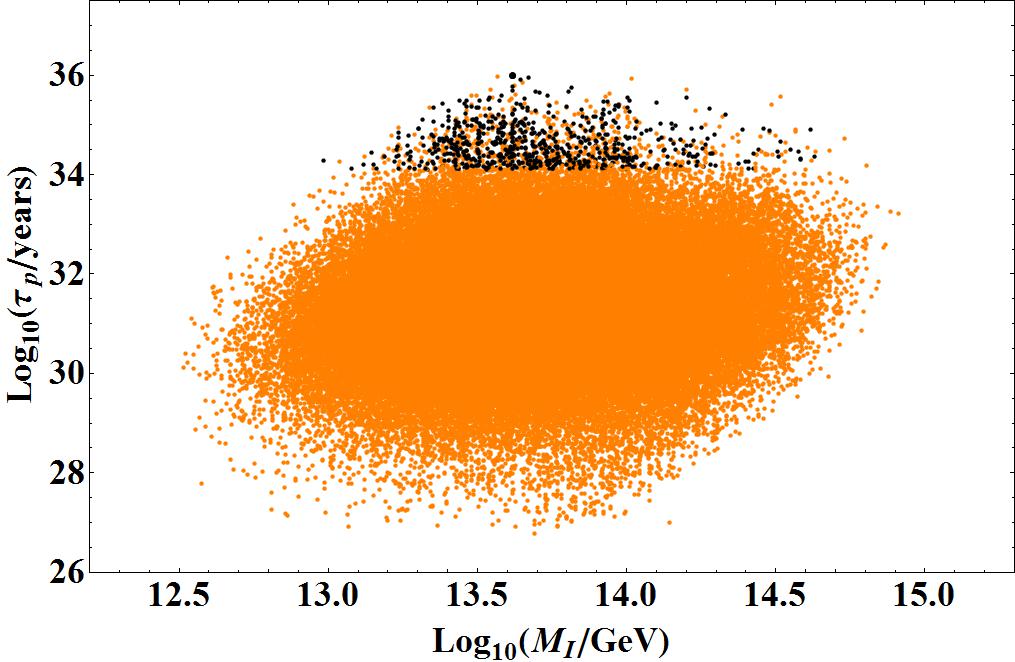}
  \caption{$R^{-1}=20$}
  \label{fig:proton lifetime vs Mps Rt=20}
\end{subfigure}
\par\bigskip
\begin{subfigure}{.5\textwidth}
  \centering
  \includegraphics[width=0.95\linewidth]{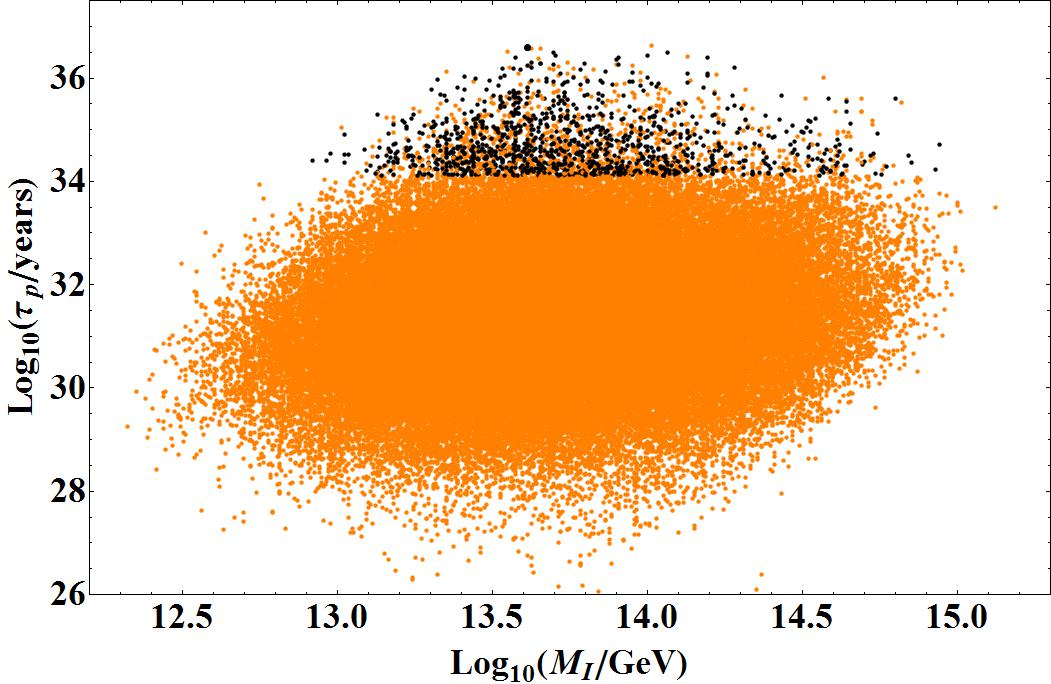}
  \caption{$R^{-1}=33$}
  \label{fig:proton lifetime vs Mps Rt=33}
\end{subfigure}
\caption{Proton lifetime $(\tau_p)$ as a function of the intermediate scale $M_I$ for different levels of threshold corrections. The ratio of the mass of each Higgs boson to the corresponding gauge boson mass is kept in between $R=\left\lbrace\nicefrac{1}{10},\nicefrac{1}{20},\nicefrac{1}{33}\right\rbrace$ and $R=2$, with $R = \frac{M_{\text{Higgs boson}}}{\text{Corresponding gauge boson mass}}$. All the black points are phenomenologically viable ones. Orange points either go through gauge boson mediated proton decay with a lifetime shorter than the experimental limit ($1.29 \times 10^{34}\;\text{yrs}$), or they correspond to scenarios where at least one of the color triplet Higgs boson acquires a mass less than $10^{13}\;\text{GeV}$. From a conservative point of view, we decided to exclude points  with light color triplet masses as they tend to contribute to proton decay at a dangerous level.}
\label{fig:proton lifetime vs Mps scrt plot}
\end{figure}
We have plotted in Fig. \ref{fig:proton lifetime vs Mps scrt plot} proton lifetime as a function of the intermediate scale $M_I$ which has been picked up as the mass of the PS gauge boson with random distribution of Higgs boson masses. The Higgs boson masses were randomly chosen among the corresponding scale and the extreme (minimum and maximum) values. All the black points correspond to phenomenologically viable points, while the orange points are excluded due to the proton lifetime experimental bound. The maximum proton lifetime in each case is marked by a larger black point. The points above the current lifetime limit $(\tau_p= 1.29 \times 10^{34} yrs)$, yet orange in the scatter plot (in Fig. \ref{fig:proton lifetime vs Mps scrt plot}) are due to the fact that those points corresponds to cases where at least one of the Higgs color triplets becomes lighter than the conservative lower limit of $10^{13}$ GeV on their masses.

\section{The $54_H + 126_H$ Higgs Model for $SO(10)$ symmetry breaking}
\label{sec:The $54_H + 126_H$ Higgs Model for $SO(10)$ symmetry breaking}
In this section we analyze the breaking of $SO(10)$ symmetry down to the SM. We consider $SO(10)$ model with Higgs sector including a real $54_H$ (a rank two symmetric tensor, denoted by $\Phi_{ij}$), a complex $126_H$ (a rank five totally antisymmteric tensor, denoted by $\Sigma_{ijklm}$) and a complex $10_H$ (a vector representation, denoted by $\phi_i$). Being non-supersymmetric, the model inherits a couple of crucial deficiencies, namely failure to address the hierarchy issue  and the lack of an automatic dark matter candidate. As the hierarchy problem is one of ``naturalness", in this model we invoke fine-tuning to bring the Higgs doublet to the weak scale and fulfill the phenomenological constraints. The loss of the dark matter candidate can be easily remedied by introducing an additional Pecci-Quinn (PQ) $U(1)_{PQ}$ symmetry to the $SO(10)$ framework. It should be stressed that in this upgraded framework, along with the dark matter candidate, one simultaneously solves the strong CP problem - the absence of CP violation in the strong interaction sector.

The assigned PQ-charges of the various fields in the model are as follows:
\begin{equation}
\label{eq:PQcharge}
\bm{16_F} \rightarrow e^{- i \alpha} \bm{16_F};\hspace{.5cm} \bm{10_H} \rightarrow e^{-2 i \alpha} \bm{10_H}; \hspace{.5cm} \bm{126_H} \rightarrow e^{2 i \alpha} \bm{126_H}; \hspace{.5cm} \bm{S_H} \rightarrow e^{-4 i \alpha}\bm{S_H} .
\end{equation}
As $\left\langle 126_H \right\rangle$ can only break a linear combination of the $U(1)_X$ and $U(1)_{PQ}$, one singlet ($S_H$) Higgs has been introduced to break the other linear combination so that only one $U(1)_Y$ is left unbroken above the weak scale \cite{PQLinearCombreak:1982}. With the newly introduced singlet field one can write the general potential for the model $SO(10) \times U(1)_{PQ}$ as
\begin{equation}
\label{Full Potential}
V=V(\Phi, \Sigma)+V(\Phi, \Sigma,\phi)+V(S)
\end{equation}
where
\begin{eqnarray}
V(\phi,\Sigma)&=&-\dfrac{\mu^2}{2} \Phi_{ij}\Phi_{ij}+ \dfrac{c}{3} \Phi_{ij} \Phi_{jk} \Phi_{ki}+ \dfrac{a}{4} \Phi_{ij}\Phi_{ij}\Phi_{kl}\Phi_{kl}+\dfrac{b}{2} \Phi_{ij}\Phi_{jk}\Phi_{kl}\Phi_{li}  - \dfrac{\nu^2}{2 \cdot 5!}\Sigma_{ijklm} \Sigma^*_{ijklm} \nonumber \\&&+\dfrac{\lambda_0}{2!^2 5!^2}\Sigma_{ijklm} \Sigma^*_{ijklm}\Sigma_{nopqr} \Sigma^*_{nopqr}  +\dfrac{\lambda_2}{4!^2}\Sigma_{ijklm} \Sigma^*_{ijkln}\Sigma_{opqrm} \Sigma^*_{opqrn} \nonumber \\&& +\dfrac{\lambda_4}{3!^2 2!^2}\Sigma_{ijklm} \Sigma^*_{ijkno}\Sigma_{pqrlm} \Sigma^*_{pqrno} +\dfrac{\lambda'_4}{3!^2}\Sigma_{ijklm} \Sigma^*_{ijkno}\Sigma_{pqrln} \Sigma^*_{pqrmo} \nonumber  \\ &&+ \dfrac{\alpha}{2! 5!}\Phi_{ij}\Phi_{ij} \Sigma_{pqrlm} \Sigma^*_{pqrlm}+\dfrac{\beta}{3!} \Phi_{ij}\Phi_{kl} \Sigma_{mnoik} \Sigma^*_{mnojl},\nonumber\\
V(\Phi,\Sigma,\phi)&=& -\xi_0^2 \phi_i \phi^*_i +\xi_1 \phi_i \phi^*_i \phi_j \phi^*_j   \nonumber + \xi_2 \phi_i \phi_i \phi^*_j \phi^*_j + \xi_3 \Phi_{i,j} \phi_i \phi^*_j  + \dfrac{\gamma_1}{4!}\Sigma_{ijklm} \Sigma^*_{ijkln} \phi_m \phi^*_n \\&&+ \dfrac{\gamma_2}{4!}\Sigma_{ijklm} \Sigma^*_{ijkln} \phi_n \phi^*_m + \dfrac{\eta_0}{2}\Phi_{ij} \Phi_{ij}\phi_k \phi^*_k  \nonumber  +  \dfrac{\eta_1}{\left(3!\right)^2\left(2!\right)^2}\Sigma_{ijklm} \Sigma^*_{ijkpq} \Sigma^*_{lmpqn} \phi_n \nonumber \\&& +  \dfrac{\eta_1^*}{\left(3!\right)^2\left(2!\right)^2}\Sigma^*_{ijklm} \Sigma_{ijkpq} \Sigma_{lmpqn} \phi^*_n    +\eta_2 \Phi_{ij} \Phi_{ik} \phi_j \phi^*_k  + \dfrac{\eta_3}{4!}\Sigma_{ijklm} \Sigma_{ijkln} \phi_m \phi_n \nonumber \\&& + \dfrac{\eta_3^*}{4!}\Sigma^*_{ijklm} \Sigma^*_{ijkln} \phi^*_m \phi^*_n, \nonumber\\
V(S)&=&
-\mu^2_s S S^* + \chi_1  \left( S S^* \right)^2 + \chi_2  \Sigma_{ijklm} \Sigma^*_{ijklm} S S^*  + \chi_3 \Phi_{ij}\Phi_{ij} S S^*  +  \dfrac{\chi_4}{4!}\Sigma_{ijklm} \Sigma_{ijkln} \Phi_{ij} S\nonumber \\&& +  \dfrac{\chi_4^*}{4!}\Sigma^*_{ijklm} \Sigma^*_{ijkln} \Phi_{ij} S^* + \chi_5 \phi_i \phi_i^* S S^* +\chi_6 \phi_i \phi_i S^* +\chi_6^* \phi_i^* \phi_i^* S. \nonumber
\end{eqnarray}
In the potential, terms like $\Sigma^4, \left(\Sigma^*\right)^4 $ are absent due to the PQ symmetry. Notice that the potential has four complex couplings, namely $\eta_1, \eta_3, \chi_4, \chi_6$. Among them, $\eta_3$ does not appear in the minimization condition and the mass spectrum. The other three couplings ($\eta_1,\chi_4, \chi_6$) can be made real by redefinitions of the fields $\Sigma, S, \phi$ respectively. This results in a vev structure of the potential ($\left\langle V \right\rangle$) which is completely devoid of any complex couplings. In short, this means that we can always find a solution where all the vev's of the fields are real and we will deal with such a case.

In the model, the cubic coupling, $c$ of $54_H^3$ is imperative for the desired symmetry breaking scenario. It has been shown in Ref. \cite{Li:1973mq} that if one tries to break $SO(10)$ with a $54_H$ in the absence of this cubic term, it only breaks it down to either $SO(5) \times SO(5)$ or $SO(9)$ .

As the $10_H$ is complex, $126_H \cdot \overline{126}_H \cdot 10_H \cdot 10_H^*$ has two linearly independent couplings ($\gamma_1, \gamma_2$). In the potential (Eq. (\ref{Full Potential})), the trivial coupling is the linear sum of the two. So, one finds that, the mass spectrum of the non-singlet Higgs only depends on the difference of the couplings~($\gamma_1 -  \gamma_2$). 

The term $126_H \cdot \overline{126}_H \cdot 126_H \cdot 10_H$ with the coefficient $\eta_1$ is important for the fermion mass fitting, as this is the term which generates the induced vev for the electroweak doublets contained in the $126_H$, or more precisely in the $(15,2,2)_{PS}$ \cite{Babu:1992ia}.
\begin{equation}
\left\langle 10_H \right\rangle \sim \eta_1 \left( \dfrac{\left\langle 126_H \right\rangle^2 }{M^2_{(15,2,2)_{PS}}} \right) \left\langle (15,2,2)_{PS} \right\rangle.
\end{equation}
\subsection{Details of Symmetry Breaking}
$SO(10)$ symmetry spontaneously breaks down to $SU(4)_C \times SU(2)_L \times SU(2)_R \times D$ when the $54_H$ acquires a non-zero vacuum expectation value given by
\begin{Large}
\begin{equation*}
 \left\langle 54 \right\rangle = diagonal \left( \nicefrac{\text{-}2}{5}, \nicefrac{\text{-}2}{5}, \nicefrac{\text{-}2}{5}, \nicefrac{\text{-}2}{5}, \nicefrac{\text{-}2}{5}, \nicefrac{\text{-}2}{5}, \nicefrac{3}{5},\nicefrac{3}{5},\nicefrac{3}{5},\nicefrac{3}{5}\right) \omega_s
\end{equation*}
\end{Large}
\noindent
\hspace{-.3cm} where $\omega_s \sim M_U$. The vev of the SM singlet from $54_H$ is $\left\langle S_{54} \right\rangle = -\sqrt{\dfrac{12}{5}} \omega_s$.
The $SU(4)_C \times SU(2)_L \times SU(2)_R \times D$ is broken down to SM model gauge group with the added $U(1)_{PQ'}$ (unbroken combination of $U(1)_X \times U(1)_{PQ}$) by the vev of $126_H$, denoted by $\left\langle \Sigma_{2,4,6,8,10} \right\rangle = \dfrac{\sigma}{4 \sqrt{2}}$, or in terms of SM Singlet in 126 as $\left\langle S_{126} \right\rangle = \dfrac{\sigma}{\sqrt{2}}$. Lastly, the singlet $S_H$ acquires a vev (denoted by $\left\langle S_S \right\rangle = \dfrac{v_s}{\sqrt{2}}$) which breaks the extra $U(1)_{PQ'}$ and we get SM plus an axion at low energy. One linear combination of the two complex $SU(2)_L$ doublets of complex $10_H$ and the two complex $SU(2)_L$ doublets of $126_H$ remains massless at this stage. This linear combination is the field that acquires a vev and breaks the electroweak symmetry. We will denote the vev's of the two complex doublets in the $10_H$ as $v_u$, $v_d$ and  the two complex doublets in the $(15,2,2)_{PS}$ in $126_H$ get the induced vev's denoted by $\kappa_u$, $\kappa_d$.

In short, the high scale vev's acquired by the SM singlet contained in the Higgs $54_H$, $126_H$ and $S_H$ are as follows:
\begin{equation}
\left\langle S_{54} \right\rangle = -\sqrt{\dfrac{12}{5}} \omega_s; \hspace{1 cm}\left\langle S_{126} \right\rangle = \dfrac{\sigma}{\sqrt{2}}; \hspace{1 cm}\left\langle S_S \right\rangle = \dfrac{v_s}{\sqrt{2}}.
\end{equation}

In this notation the vacuum expectation value of the potential ($V$) becomes:
\begin{align}
\left\langle V \right\rangle = & -\dfrac{6}{5} \mu^2 \omega_s^2+\dfrac{4}{25} c \omega_s^3 +\dfrac{36}{25} a \omega_s^4 +\dfrac{42}{125} b \omega_s^4 - \dfrac{\nu^2}{2}\sigma^2 +\lambda_0 \sigma^4  +\dfrac{3}{5} \alpha \omega_s^2 \sigma^2 -\dfrac{3}{5} \beta \omega_s^2   \sigma^2 \nonumber\\&-\dfrac{\mu^2_s}{2} v_s^2 +\dfrac{\chi_1}{4} v_s^4 +\dfrac{\chi_2}{4} \sigma^2 v_s^2 + \dfrac{3}{5} \chi_3 \omega_s^2 v_s^2 .
\end{align}
Here the weak scale vev's are ignored as they are atleast $10^{-8}$ times smaller than the smallest vev (namely $v_s$). It is very much possible to keep the electroweak scale vev's in the equations and do all the corresponding calculations. At the end of the calculation, one will realize that the weak scale vev's will only correspond to the mass splitting of the electroweak multiplets and that will correspond to an order of $10^2\;\text{GeV}$. Besides the SM doublet, all other scalar fields will acquire masses at the order of $10^{10}\;\text{GeV}$ or higher. So, for the sake of simplicity, we shall ignore corrections of order of $10^2 \;\text{GeV}$, which is well justified.

Minimizing $\left\langle V \right\rangle$ with respect to the parameters $\omega_s, \sigma, v_s$, the following relations can be obtained:
\begin{align}
\mu^2 = & \dfrac{12}{5} a \omega_s^2 +\dfrac{14}{25} b \omega_s^2 +\dfrac{c}{5} \omega_s+\dfrac{\alpha}{2} \sigma^2 -\dfrac{\beta}{2} \sigma^2+\dfrac{\chi_3}{2} v_s^2 \nonumber \\
\nu^2 = & \dfrac{6}{5} \alpha \omega_s^2 -\dfrac{6}{5} \beta \omega_s^2 + \lambda_0 \sigma^2 \dfrac{\chi_2}{2} v_s \nonumber \\
\mu_s^2 = & \chi_1 v_s + \dfrac{\chi_2}{2} \sigma^2 +\dfrac{6}{5} \chi_3 \omega_s^2 .
\end{align}
These minimization conditions are used in determining the masses of the various Higgs multiplets in the next subsection.
\subsection{Tree level mass spectrum}
One can go ahead and analyze the potential in its full glory (for example, using the methods described in Ref. \cite{aulakh}) and determine the whole scalar mass spectrum. The gauge boson mass spectrum is determined by constructing the covariant derivative and then analyzing the kinetic part of the Lagrangian, as usual.
\subsubsection*{Gauge boson mass spectrum}
One needs to consider the properly normalized fields and analyze the kinetic part of the Lagrangian to obtain the mass spectrum of all the gauge bosons. In the case of the gauge bosons, besides the usual leptoquark gauge bosons $(X,Y)(3,2,\nicefrac{\text{-}5}{6})$, $(X',Y')(3,2,\nicefrac{\text{+}1}{6})$ and the Pati-Salam gauge boson $(\overline{3},1,\nicefrac{\text{-}2}{3})$ and their conjugates, the heavy $(1,1,\pm1)$ particle corresponds to the right handed $W_R^\pm$. One of the $(1,1,0)$ corresponds to the $Z'$ of the $U(1)_R$ and other one to the weak scale $Z$-boson which remains massless until the electroweak symmetry is broken. We find these masses to be:
\begin{align}
\label{eq:gaugebosonspectrum}
M_A^2(3,2,-\dfrac{5}{6})=& g^2 \omega_s^2;\nonumber\\
M_A^2(3,2,+\dfrac{1}{6})=& g^2 (\omega_s^2 + \sigma^2);\nonumber\\
M_A^2(\overline{3},1,-\dfrac{2}{3})=& g^2 \sigma^2;\nonumber\\
M_A^2(1,1,-1)=& g^2 \sigma^2; \nonumber\\
M_A^2(1,1,0)=&\left( \begin{array}{cc}
3 & \sqrt{6} \\
\sqrt{6} & 2 \end{array} \right) g^2 \sigma^2 .
\end{align}
Here the quantum numbers listed are under SM group $SU(3)_C \times SU(2)_L \times U(1)_Y$. One should notice that the determinant of the mass matrix for the gauge boson $(1,1,0)$ is zero and the eigenvalues of the matrix are  $\left\lbrace 5,0\right\rbrace$. The zero eigenvalue corresponds to the $Z$-boson of mass $91$ GeV.
\subsubsection*{Scalar Boson Mass Spectrum}
The determination of the scalar mass spectrum is a little bit involved compared to  the gauge boson mass spectrum, mainly due to presence $126_H$ which is represented by a rank-five totally antisymmetric tensor. But as shown in Ref. \cite{aulakh}, one can identify the sub-multiplets inside the full multiplet in a systematic way and insert the vev's for the SM singlets to obtain the scalar mass spectrum.

Going through the straightforward, yet tedious calculation, one gets the following mass spectrum:
\begin{align}
& M^2(1,3,0)= \frac{8}{5} b \omega_s^2 + c \omega_s; \nonumber \\
& M^2(8,1,0)= \frac{2}{5} b \omega_s^2 - c \omega_s; \nonumber \\
& M^2(3,3,-\frac{1}{3}) = 4 \left( 3 \lambda_2 + 3 \lambda_4 + 4 \lambda'_4\right) \sigma ^2; \nonumber \\
& M^2(6,3,+\frac{1}{3}) = 8 \left(  \lambda_2 +  \lambda_4 + 4 \lambda'_4\right) \sigma ^2; \nonumber \\
&M^2(3,2,-\frac{5}{6})=0;\nonumber\\
&M^2(\overline{3},1,-\frac{2}{3})=0;\nonumber\\
&M^2(1,1,-1)=0\nonumber
\end{align}
\begin{align}
& M^2(1,1,+2) = 8 \left(  \lambda_2 +  \lambda_4 + 4 \lambda'_4\right) \sigma ^2; \nonumber \\
& M^2(\overline{3},1,+\frac{4}{3}) = 4 \left( 3 \lambda_2 + 3 \lambda_4 + 4 \lambda'_4\right) \sigma ^2; \nonumber \\
& M^2(\overline{6},1,-\frac{4}{3}) = 8 \left(  \lambda_2 +  \lambda_4 + 4 \lambda'_4\right) \sigma ^2; \nonumber \\
& M^2(\overline{6},1,-\frac{1}{3}) = 4 \left( 3 \lambda_2 + 3 \lambda_4 + 4 \lambda'_4\right) \sigma ^2; \nonumber \\
& M^2(1,3,-1)=\left( \begin{array}{cc}
\frac{8}{5} b \omega_s^2 + c \omega_s+\frac{1}{2} \beta \sigma ^2 &   2  \chi_4 \sigma v_s \\
2 \chi_4 \sigma v_s & 8 \left( 2 \lambda_2 + 3 \lambda_4 + 2 \lambda'_4 \right) \sigma ^2
\end{array} \right); \nonumber\\
& M^2(\overline{6},1,+\frac{2}{3})=\left( \begin{array}{cc}
\frac{2}{5} b \omega_s^2 - c \omega_s+\frac{1}{2} \beta \sigma ^2 &  2  \chi_4 \sigma v_s \\
2  \chi_4 \sigma v_s & 8 \left( 2 \lambda_2 + 3 \lambda_4 + 2 \lambda'_4\right) \sigma ^2
\end{array} \right); \nonumber\\
& M^2(3,2,+\frac{7}{6})=\left( \begin{array}{cc}
4 \left( 3 \lambda_2 + 3 \lambda_4 + 4 \lambda'_4\right) \sigma ^2 + \beta \omega_s^2 & -2 \sqrt{2} \chi_4 \omega_s v_s \\
-2 \sqrt{2} \chi_4 \omega_s v_s & 8 \left(  \lambda_2 +  \lambda_4 + 2 \lambda'_4\right) \sigma ^2 + \beta \omega_s^2
\end{array} \right); \nonumber \\
& M^2(8,2,-\frac{1}{2})=\left( \begin{array}{cc}
4 \left( 3 \lambda_2 + 3 \lambda_4 + 4 \lambda'_4\right) \sigma ^2 + \beta \omega_s^2 & -2 \sqrt{2} \chi_4 \omega_s v_s \\
- 2 \sqrt{2} \chi_4 \omega_s v_s & 8 \left(  \lambda_2 +  \lambda_4 + 2 \lambda'_4\right) \sigma ^2 + \beta \omega_s^2
\end{array} \right); \nonumber\\
& M^2(3,2,+\frac{1}{6})=\left( \begin{array}{ccc}
 8 \left( 2 \lambda_2 + 3 \lambda_4 + 2 \lambda'_4\right) \sigma ^2 +\beta \omega_s^2 & 2 \sqrt{2} \chi_4 \omega_s v_s & -2 \chi_4 \sigma v_s\\
2 \sqrt{2} \chi_4 \omega_s v_s & \beta \omega_s^2 & \frac{1}{\sqrt{2}} \beta \sigma \omega_s \\
- 2 \chi_4 \sigma v_s &  \frac{1}{\sqrt{2}} \beta \sigma \omega_s & \frac{1}{2} \beta  \sigma ^2
\end{array} \right); \nonumber \\
& M^2(3,1,-\frac{1}{3})=\left( \scalemath{0.8}{\begin{array}{ccccc}
4 \left( 3 \lambda_2 + 3 \lambda_4 + 4 \lambda'_4 \right) \sigma ^2 + 4 \beta \omega_s^2 & 4\sqrt{2} \chi_4 \omega_s v_s & 0 & 0 & 0 \\
4 \sqrt{2} \chi_4 \omega_s v_s & 8 \left(  \lambda_2 +  \lambda_4 + 2 \lambda'_4 \right) \sigma ^2 + 4 \beta \omega_s^2 & 16  \sqrt{2} \lambda'_4 \sigma ^2 & 0 & 4 \eta_1 \sigma ^2 \\
0 & 16  \sqrt{2} \lambda'_4 \sigma ^2 & 8 \left(  \lambda_2 +  \lambda_4 \right)\sigma ^2 & 0 &  4 \sqrt{2} \eta_1  \sigma^2  \\
0& 0&0  & A1 & \sqrt{2} \chi_6 v_s \\
0 & 4 \eta_1 \sigma ^2 & 4 \sqrt{2} \eta_1  \sigma^2 & \sqrt{2} \chi_6 v_s & B1 
\end{array}} \right); \nonumber \\
&M^2(1,2,-\frac{1}{2})=\left( \scalemath{0.8}{\begin{array}{cccc}
 8 \left(  \lambda_2 +  \lambda_4 - 2 \lambda'_4\right) \sigma ^2 +\beta \omega_s^2 & 2 \sqrt{2} \chi_4 \omega_s v_s & 0& 4  \sqrt{3}\eta_1 \sigma ^2 \\
2\sqrt{2} \chi_4 \omega_s v_s & 4 \left( 3 \lambda_2 + 3 \lambda_4 +4 \lambda'_4 \right) \sigma ^2 +\beta \omega_s^2 & 0 &0 \\
 0  & 0 & A2 & \sqrt{2} \chi_6 v_s \\ 
4  \sqrt{3} \eta_1 \sigma ^2 & 0 & \sqrt{2} \chi_6 v_s & B2
\end{array}} \right); \nonumber \\
& M^2(1,1,0)= \left( \scalemath{0.8}{\begin{array}{ccc}
\frac{1}{10} c \omega_s +\frac{12}{5} a \omega_s^2+\frac{14}{25}b \omega_s^2 & - \sqrt{\frac{3}{5}} (\alpha-\beta) \sigma \omega_s & -\sqrt{\frac{3}{5}} \chi_3 \omega_s v_s \\
 -\sqrt{\frac{3}{5}} (\alpha-\beta) \sigma \omega_s & \frac{1}{4} \lambda_0 \sigma^2  &  \frac{1}{2}\chi_2 \sigma v_s \\
-\sqrt{\frac{3}{5}} \chi_3 \omega_s v_s &  \frac{1}{2} \chi_2 \sigma v_s &  \chi_1 v_s^2
\end{array}} \right). \nonumber \\
&M^2(1,1,0) =0\nonumber\\
&M^2(1,1,0) =0
\end{align}
Here in the color triplet and $SU(2)_L$ doublet matrices, we have defined
\begin{align}
\label{\eq:HiggsSpectrum}
A1=& -\frac{2}{5}\xi_3 \omega_s+\frac{6}{5}\eta_0 \omega_s^2+\frac{4}{25}\eta_2 \omega_s^2+\gamma_1 \sigma^2+\frac{1}{2}\chi_5 v_s^2+ m^2\nonumber \\
B1=& -\frac{2}{5}\xi_3 \omega_s+\frac{6}{5}\eta_0 \omega_s^2+\frac{4}{25}\eta_2 \omega_s^2+\gamma_2 \sigma^2+\frac{1}{2}\chi_5 v_s^2+ m^2\nonumber \\
A2=& \frac{3}{5}\xi_3 \omega_s+\frac{6}{5}\eta_0 \omega_s^2+\frac{9}{25}\eta_2 \omega_s^2+\gamma_1 \sigma^2+\frac{1}{2}\chi_5 v_s^2 + m^2\nonumber \\
B2=& \frac{3}{5}\xi_3 \omega_s+\frac{6}{5}\eta_0 \omega_s^2+\frac{9}{25}\eta_2 \omega_s^2+\gamma_2 \sigma^2+\frac{1}{2}\chi_5 v_s^2 + m^2\nonumber 
\end{align}
The mass matrices are spanned in the following bases:
\begin{align}
(1,3,-1) & \rightarrow \left\lbrace (1,3,-1)_{\zeta_1},(1,3,-1)_{\Sigma_{21}}\right\rbrace \nonumber \\
(\overline{6}, 1, -\nicefrac{2}{3}) & \rightarrow \left\lbrace (\overline{6}, 1, -\nicefrac{2}{3})_{\zeta_3}, (\overline{6}, 1, -\nicefrac{2}{3}))_{\Sigma_{39}} \right\rbrace \nonumber \\
(3,2,+\nicefrac{7}{6}) & \rightarrow \left\lbrace (3,2,+\nicefrac{7}{6})_{\Sigma^*_{46}},(3,2,+\nicefrac{7}{6})_{\Sigma_{43}} \right\rbrace \nonumber \\
(8,1,-\nicefrac{1}{2}) & \rightarrow \left\lbrace (8,1,-\nicefrac{1}{2})_{\Sigma^*_{48}},(8,1,-\nicefrac{1}{2})_{\Sigma_{47}} \right\rbrace \nonumber \\
(3,2,+\nicefrac{1}{6}) & \rightarrow \left\lbrace (3,2,+\nicefrac{1}{6})_{\Sigma^*_{45}},(3,2,+\nicefrac{1}{6})_{\Sigma_{44}},(3,2,+\nicefrac{1}{6})_{\zeta_2} \right\rbrace\nonumber \\
(3,1,-\nicefrac{1}{3}) & \rightarrow \left\lbrace (3,1,-\nicefrac{1}{3})_{\Sigma_{11}},(3,1,-\nicefrac{1}{3})_{\Sigma^*_{12}},(3,1,-\nicefrac{1}{3})_{\Sigma^*_{35}},(3,1,-\nicefrac{1}{3})_{T_1},(3,1,-\nicefrac{1}{3})_{T^*_2} \right\rbrace\nonumber \\
(1,2,-\nicefrac{1}{2}) & \rightarrow \left\lbrace (1,2,-\nicefrac{1}{2})_{\Sigma_{41}},(1,2,-\nicefrac{1}{2})_{\Sigma^*_{42}},(1,2,-\nicefrac{1}{2})_{H_1},(1,2,-\nicefrac{1}{2})_{H^*_2} \right\rbrace \nonumber
\end{align}

A few remarks are in order about the mass spectrum:
\begin{itemize}
\vspace{-.6cm}
\item From the mass spectrum of the Higgs bosons, we find that there are 34 massless states, which correspond to the broken generators of $SO(10)$ plus the imaginary part of the singlet $(S)$ which corresponds to the axion. These 33 Goldstone bosons are eaten up by the 33 massive gauge bosons whose mass spectrum is given in Eq. (\ref{eq:gaugebosonspectrum}).
\item From the mass spectrum, it is obvious that $\Sigma_{22}(3,3,-\nicefrac{1}{3})$, $\Sigma_{38}(\overline{6}, 1, -\nicefrac{1}{3})$, $\Sigma_{42}(1,2,+\nicefrac{1}{2})$, $\Sigma_{46}(\overline{3},2,-\nicefrac{7}{6})$, $\Sigma_{48}(8,2,+\nicefrac{1}{2})$ and $\Sigma_{11}(3,1, -\nicefrac{1}{3})$ are degenerate except for the presence of contribution coming $54_H$ vev ($\omega_s$) for the Higgs $\Sigma_{42},\Sigma_{46},\Sigma_{48}$ and $\Sigma_{11}$. The degeneracy comes from the fact that all these Higgs bosons are inside the $(45,2)$under $SU(5)\times  U(1)_X$. As $\Sigma_3(\overline{10},1,3)$ acquires a vev, minimization condition removes the $\omega_s$ contribution from $\Sigma_{38} $ and due to the $D$-parity, $\Sigma_{22}$ is also missing the $\omega_s$ contribution to its mass.
\item Similar arguments apply for masses of $\; \Sigma_{23}(6,3,+\nicefrac{1}{3})$, $\; \Sigma_{33}(1,1,+2)$, $\; \Sigma_{35}(\overline{3}, 1, +\nicefrac{1}{3})$,   $\Sigma_{37}(\overline{6}, 1, -\nicefrac{4}{3})$, $\Sigma_{43}(3,2,+\nicefrac{7}{6})$ and $\Sigma_{47}(8,2,-\nicefrac{1}{2})$ where only $\Sigma_{43}$ and $\Sigma_{47}$ are the only ones having contribution from $\omega_s$.
\item The rank of the matrix $M^2(3,2, +\nicefrac{1}{6})$ is two, where the massless eigenstate corresponds to the Goldstone boson of the theory, which gets eaten up by the $(X',Y')$ gauge boson. The other massless Goldstone bosons from $(3,2,-\nicefrac{5}{6})$ are absorbed by the $(X,Y)$ gauge boson.
\item In the absence of the singlet ($S_H$), most of the mass matrices reduce to diagonal forms, indicating no mixing between many fields, even though they possess the same gauge quantum numbers. This is again due to the fact that the corresponding SM fields reside in a different $SU(5)$ multiplet. For example, one of the $(8,2,-\nicefrac{1}{2})$ lives in the $(\overline{50},+2)$ and a second one is in the $(45,-2)$ under $SU(5) \times U(1)_X$. There is non-trivial mixing in the mass matrix of $(3,2, +\nicefrac{1}{6})$ due to non-trivial quartic coupling $\beta$ which generates term like $(24,0)(1,+10)(15,-4)(\overline{15},-6)$ written under $SU(5)\times U(1)_X$.
\item In the doublet mass matrix only one of the doublets from $126_H$ gets mixed up directly (due to the coupling $\eta_1$) with one of the doublets from $10_H$-plet as they both are from $(\overline{5},+2)$ of $SU(5)\times U(1)_X$. Besides this term, the two Higgs doublets in $126_H$ get mixed due to the presence of the couplings from the singlet-potential ($V(S)$). Same type of mixing happens between the two Higgs doublet in $10_H$. So, all the doublet fields mix with each other. This property of the doublet mass matrix is of utmost importance to generate realistic fermion masses.
\item One should remember that in $SO(10)$ models without the PQ-symmetry, due to the presence of the term $126 \cdot 126 \cdot 54$ in the Lagrangian, one will end up getting all the off-diagonal mixing term at the SM level Lagrangian. Even though one gets a well-mixed doublet mass matrix, in that case one ends up with an extra set of Yukawa couplings and the theory loses predictivity. The inclusion of the PQ-symmetry gets rid of the extra Yukawa couplings and at the same time, gets rid of the usual mixing terms in the scalar mass spectra. But, at the end, the couplings in the singlet part of the potential ($V(S)$), which breaks the PQ-symmtery, reintroduces those mixing terms in the mass matrices. This makes the singlet vev important as in the doublet mass matrix it shows up in the off-diagonal terms and in the Yukawa sector the off diagonal elements cannot be ignored while reproducing realistic fermion masses and mixings.
\item As one of the electroweak doublets and the color antitriplets of $126_H$ live in $(\overline{5},+2)$ of $SU(5) \times U(1)_X$ and one of the electroweak doublets and color triplets of $10_H$ live in $(5,2)$, the mixing term in doublet mass matrix and triplet mass matrix should be the same ($4 \sqrt{3} \eta_1 \sigma^2$). The apparent dissimilarity in the triplet matrix is due to the basis in which the triplet mass matrix is written. By a simple rotation of the triplet mass matrix one can show that mixing term is indeed given by $4 \sqrt{3} \eta_1 \sigma^2$ and in that basis, there is no mixing between the triplet from $(\overline{50},+2)$ and triplet from $10_H$-plet.  Besides these ones, other terms in the doublet and triplet matrices differ in a significant way and fine-tuning the doublet determinant to zero does not set the determinant of the triplet mass matrix to zero. While not unexpected, this condition is crucial for consistent phenomenology.
\item The presence of the quartic coupling $\beta$ ensures a mass contribution from the $\left\langle 126_H \right\rangle$ for the fields from $54_H$. But such a contribution is missing from the multiplet which resides in the $24_H$-plet of $SU(5)$. This can be easily explained under a situation where the $\left\langle 126_H \right\rangle >\left\langle 54_H \right\rangle$ and the model goes through $SU(5)$. Now that the SM singlet of $54_H$ is in the $24_H$-plet (under $SU(5)$) and the minimization condition removes the contribution from $\left\langle 126_H \right\rangle$, SM muliplet $(8,1,0)$ and $(1,3,0)$ end up with no mass contribution from $\left\langle 126_H \right\rangle$. For the same reason only $(X',Y')$ gauge bosons have mass contribution from the $\left\langle 126_H \right\rangle$, but not $(X,Y)$ gauge bosons.
\end{itemize}
\section{Yukawa sector of the model}
\label{sec:Yukawa sector of the model}
The burning question about the search of the minimal Yukawa sector can be addressed under this minimal $SO(10)$ model with PQ-symmetry. In $SO(10)$ grand unified theory, each generation of fermions belong to a 16-dimensional spinorial representation, whose masses arise from the renormalizable Yukawa couplings with Higgs fields $(\overline{16} \times \overline{16}= 10 + \overline{126} + 120)$. In the minimal model described here, the Yukawa part of the Lagrangian is given as:
\begin{equation}
\mathcal{L}=16_F \left( Y_{10} 10_H +Y_{\overline{126}} \overline{126}_H\right) 16_F
\end{equation} 
where $Y_{10}, Y_{\overline{126}}$ are complex symmetric matrices in the generation space. A complex $10_H$ in general brings an extra set of Yukawa couplings. But in this case $U(1)_{PQ}$ symmetry forbids $16_F 10^*_H 16_F$ couplings, see Eq. (\ref{eq:PQcharge}). Besides providing a candidate for dark matter while solving the strong CP problem, the PQ-symmetry also affects the Yukawa sector making it realistic and predictive \cite{Bajc:2005zf}. Notice that here, both $10_H (\phi)$ and $126_H(\Sigma)$ are complex in nature and each of them carries two $SU(2)_L$ Higgs doublets. It is assumed that only one linear combination of these electroweak doublets remains massless before electroweak symmetry breaking and acquires electroweak vev. This corresponds to the minimal fine-tuning as dictated by extended survival hypothesis. For such a case, the quark and lepton mass matrices become:
\begin{align}
M_u= h v_u + f \kappa_u; \hspace{1cm}&\hspace{1cm} M_d=h v_d + f \kappa_d; \nonumber \\
M_\nu^D= h v_u -3 f \kappa_u; \hspace{1cm}&\hspace{1cm} M_l=h v_d -3 f \kappa_d; \nonumber\\
M_\nu^M &= f \sigma. 
\end{align}
Here, $M_{u,d,l}$ is the up-type quark, down-type quark and lepton mass matrix, $M_\nu^D$ is the Dirac neutrino matrix and $M_\nu^M$ is the Majorana mass matrix. These expressions can be rewritten in a more compact form which is more popular for a fit to masses and mixing angles:
\begin{align}
M_u= r(H+s F); \hspace{1cm}&\hspace{1cm} M_d= H+ F; \nonumber \\
M_\nu^D= r(H-3 s F); \hspace{1cm}&\hspace{1cm} M_l=H-3 F; \nonumber \\
M_\nu^M &= r_R^{-1} F
\end{align}
where $H=h v_d, \; F = f \kappa_d$ are complex symmetric matrices and $r=\nicefrac{v_u}{v_d},\; s=\nicefrac{\kappa_u}{(r \kappa_d)}, \;r_R =\nicefrac{\kappa_d}{\sigma}$ are dimonsionless parameter. 

An ample amount of literature has been devoted to find the best fit values of the parameters for various general minimal and non-minimal SO(10) Yukawa structures \cite{Bajc:2001fe, Fukuyama:2002ch, Bajc:2002iw, Goh:2003sy, Goh:2003hf, Bertolini:2004eq, Babu:2005ia, Joshipura:2011nn, Altarelli:2013aqa, Dueck:2013gca}. Then it is suffice to say that the minimal model described in this work can reproduce a realistic fermion mass spectrum, if the model has enough freedom to have $r\approx 69$ and $s\approx 0.36 - 0.04 i$ with some uncertainty \cite{Joshipura:2011nn}. Here, we can see that $s$ is almost real, exactly what is needed for this model. At this point, we do realize that a small deviation in the parameter of a delicate $\chi^2$ - analysis used in the fit of fermion masses and mixings has the potential to make the $\chi^2_{min}$ shift. Under such scenario, one can always redo the $\chi^2$-analysis and minimize the $\chi^2$. Besides adjusting the input mass matrices (for example, lepton and down-type quark mass matrices $M_l,M_d$), the process has the potential to change the vev ratios ($r,s$) as in this minimal model there is no phase associated with $s$. Yet as the phase of the $s$ parameter is already small, we do not expect a large change in the fitting of fermion masses and mixings and we also emphasis the fact that the model has enough freedom to accommodate such a change.

So, one needs to verify and make sure that the doublet mass matrix has enough freedom to remain positive-definite and produce the appropriate vev ratios while not leading to light eigenvalues in the triplet mass matrix.

From the structure of the doublet mass matrix ($\mathcal{D}$), we see that the massless SM Higgs doublet $h_{SM}$ becomes
\begin{equation}
h_{SM} = \alpha_H H_u +\beta_H H_d^*+\alpha_h h_u +\beta_h h_d^*
\end{equation}
where
$H_u$ and $H_d$ are the up-type and down-type doublet in $(15,2,2)_{PS}$ of $126_H$ and $h_u$ and $h_d$ are the up-type and down-type doublet living in the complex $10_H$. For such a case we have
\begin{align}
\label{Doublet Yukawa conditions}
&\mathcal{D}_{11} \alpha_H +\mathcal{D}_{12} \beta_H + \mathcal{D}_{14} \beta_d =0; \nonumber \\
&\mathcal{D}_{12} \alpha_H + \mathcal{D}_{22} \beta_H = 0; \nonumber \\
&\mathcal{D}_{33} \alpha_h + \mathcal{D}_{34} \beta_h =0; \nonumber \\
&\mathcal{D}_{14} \alpha_H +\mathcal{D}_{34} \alpha_h +\mathcal{D}_{44} \beta_h =0; \nonumber \\
&\mathcal{D}_{11}\mathcal{D}_{22}\mathcal{D}_{33}\mathcal{D}_{44}+ \mathcal{D}_{12}^2 \mathcal{D}_{34}^2- \mathcal{D}_{22} \mathcal{D}_{33} \mathcal{D}_{14}^2-\mathcal{D}{33} \mathcal{D}_{44}\mathcal{D}_{12}^2 - \mathcal{D}_{11}\mathcal{D}_{22} \mathcal{D}_{34}^2 =0.
\end{align}
As $r=\nicefrac{\alpha_h}{\beta_h}>0$ and $\mathcal{D}_{22}>0$, we must have $\mathcal{D}_{12}<0$. Similar argument implies $\mathcal{D}_{34}<0$. Again one can show from the above mentioned set of equations that 
\begin{equation}
\mathcal{D}_{11} = \dfrac{\mathcal{D}_{14}^2}{\mathcal{D}_{44}-r^2 \mathcal{D}_{33}}-\dfrac{\mathcal{D}_{12}}{s}>0
\end{equation}
Now, without any loss of generality, one can take the sign of $\beta_h$ to be positive, then $\alpha_h>0$ and $\alpha_H$ and $\beta_H$ are of the same sign. For the case, $\alpha_H>0$, if $\mathcal{D}_{14} >0$ then only valid solution lies for $\left|\dfrac{\mathcal{D}_{14}^2}{\mathcal{D}_{44}-r^2 \mathcal{D}_{33}} \right|<\left| \dfrac{\mathcal{D}_{12}}{s} \right|$. In contrast, if $\mathcal{D}_{14} <0$, there is no such constraint. Similarly, for the case $\alpha_H<0$, the case $\mathcal{D}_{14} <0$ gets the added condition. These conditions reduce the parameter space of the model significantly and need to be considered when one starts the process of random selection of sample points for the Higgs cubic and quartic couplings.
\section{Technical Details}
\label{sec:Technical Details}
Due to the richness in the mass matrices in the scalar mass spectrum, one fails to come up with simple mass relations for the Higgs sector. All the couplings coming from the $SO(10)$-potential (Eq. (\ref{Full Potential})) need to be in the perturbative range. So, it is obvious that instead of the scalar masses, one should start from the couplings and vev's of the theory and calculate the mass spectrum of scalars and gauge bosons. Now, for a realistic model one needs to take into account the unification of the couplings and all the phenomenological constraints imposed by proton decay, realistic fermion mass spectrum and dark matter abundance.

To produce a sample scalar mass spectrum for the Higgs ($10_H$, $54_H$, $126_H$, $S_H$), first the vev's were picked to be $\omega_s \sim \left(10^{15}- 10^{16} \right)$ GeV, $\sigma \sim \left(10^{13}- 10^{14} \right)$GeV, $v_s \sim \left(10^{10}- 10^{12} \right)$GeV. The range of the intermediate scale and unification scale vev's are decided from the scatter plot generated before the scalar mass spectrum was determined (see Fig.\ref{fig:proton lifetime vs Mps scrt plot}). As  the singlet vev ($v_s$) breaks the PQ-symmetry, it corresponds to the axion decay constant $f_a$ and the range taken for $v_s$ is compatible with all the axion search experiments and astrophysical bounds. In the numerical analysis, one also sees that the above-mentioned range for $v_s$ is also preferred by the doublet mass matrix. The dimensionless couplings are chosen randomly in the range of $[-1,1]$ with the exception of the coupling $b$ which was chosen from $[-2,2]$ due to the poor availability of realistic parameter space in the more restrictive range. The couplings with positive mass dimensions were chosen either to be close to the corresponding scale or lower than the scale, so that the potential does not develop any unwanted minimum.

With the scalar masses fixed, the gauge boson masses were determined by the unification constraints and using the RGE the unified coupling at the unification scale was calculated. This unified coupling and the pre-assumed vev's also give the gauge boson masses which generally do not coincide with the previous ones determined from the unification conditions. To solve this an iteration process was used, until the difference between gauge boson masses calculated from these two methods becomes negligible.

The running of the gauge couplings can be done mainly in three ways:
\begin{enumerate}
\item One can run the SM gauge couplings at one-loop level, all the way to the unification scale while updating the beta function coefficients whenever one encounters a scalar or gauge boson. The uncertainty coming from the one-loop running can be reduced if we run the SM gauge couplings at the two-loop level until we introduce the heavy scalar particles. The full two-loop running for computing the threshold corrections is not done due to the unknown two-loop connecting formula at the scale of symmetry breaking.
\item One can do a one-loop running of the SM gauge couplings until one hits a energy scale corresponding to the Pati-Salam gauge boson $(\overline{3},1,-\nicefrac{2}{3})$. Beyond that energy scale, it is the gauge couplings of the Pati-Salam model that is considered to be evolving until we reach a energy scale corresponding to the leptoquark gauge boson $(X,Y)$. Under such type of running, unification is achieved after we have crossed the threshold of all the scalars and gauge multiplets of the theory. This program introduces some uncertainty due to the mass splittings of the sub-multiplets due to the lower order vev's. This may become important due to the vicinity of the intermediate and unification scales. To remedy the issue, while running the gauge couplings of higher symmetry, we only introduce the effects of a scalar particle in the beta functions, if that particle completes the multiplet of the higher symmetry. Finally, the uncertainty coming from the one-loop running can be reduced if we run the SM gauge couplings at the two-loop level until we introduce the first heavy scalar particles. Again, the full two-loop running was not done due to the unknown two-loop connecting formula of symmetry breaking.
\item One can also do a two-loop running where all the threshold correction is dumped in the intermediate and unification scales. Then one ends up with a discontinuity of the running of couplings corresponding to the threshold corrections. One should remember that in this case the scale at which the couplings become unified does not necessarily correspond to the mass of the leptoquark gauge bosons which mediate proton decay. One can chose the scale to be the intermediate scale and unification scale determined initially without any threshold corrections. In the following part of the paper, we picked the vev's as the corresponding scales. 
\end{enumerate}
The following steps were taken to produce the benchmark points:
\begin{itemize}
\item To produce the initial results, a set of random numbers (within a reasonable range) was generated in the 24-dimensional parameter space. Using the mass spectrum, all the Higgs boson masses were calculated and the gauge boson masses were determined by Eqs. (\ref{eq:Mps_def}), (\ref{eq:Unification scale}). One-loop running of the SM gauge couplings was performed to determine the initial status of the unification. At this level, strict unification is not achieved and the data set does not reproduce a realistic fermion mass spectrum. Also the gauge boson masses do not necessarily correspond to the one calculated from gauge boson mass spectrum. Each of those points is selected individually and updated so that all the points satisfy the consistency checks and phenomenological constraints. 
\item After the initial random choice of parameters, one needs to impose the constraints imposed by the doublet mass matrix and realistic fermion mass spectrum. One needs to update the initial parameter values to generate one massless Higgs doublet and keep the vev ratios fixed to at $r\approx 69, s\approx 0.36$ \cite{Joshipura:2011nn}. 
\item By performing a gauge coupling evolution, the gauge boson masses are updated so that we achieve $SU(2)_L$ and $SU(2)_R$ unification at Pati-Salam scale and perfect unification at the GUT scale. As the vev $\omega_s$ corresponding to the gauge boson mass and the one corresponding to scalar masses do not necessary coincide, an iteration process was run to rectify the situation. 
\item Due to the iteration process, the vev's of the theory get updated. As the doublet mass matrix, which is required to satisfy multiple conditions, is highly sensitive to the vev's, one needs to update the parameter space to ensure that availability of the massless Higgs doublet and keep the vev ratios fixed.
\item This update of parameter space requires update of the vev's by iteration process described earlier so that gauge boson masses remain consistent. These updates of vev's and coupling parameters need to be iterated until the error is within an acceptable limit.
\item At every step one also has to keep checking the positive-definiteness of all the eigenvalues of all mass matrices and pay special attention to the triplet mass matrix so that the lowest lying color triplet does not become much lighter than $10^{13} \; \text{GeV}$.
\end{itemize}
\section{Results with Benchmark points}
\label{sec:Results with Benchmark points}
In this section we present our procedure to pick a couple of benchmark points. Going through the procedure and constraints described in the previous section, we can identify sample points satisfying all the phenomenological constraints which would then become true candidates from the large parameter space. For that purpose, one can ease the process by including the conditions required to ensure the stability of the vacuum and positive-definiteness of all the scalar masses along with the issue of realistic fermion mass spectrum. 

For example: being the only non-trivial coupling of $54^2\cdot 126\cdot \overline{126}$, $\beta$ needs to be fine-tuned so that $(15,2,2)_{PS}$ stays in the intermediate scale. The conditions translates as $\beta \omega_s^2 \approx \sigma^2$, making $\beta \lesssim \nicefrac{\sigma^2}{\omega_s^2}\approx 10^{-4}$, while positive-definiteness of the mass matrix of $(3,2, + \nicefrac{1}{6})$ says $\beta >0$. Again, from the masses of $(1,3,0)$ and $(8,1,0)$, we can say that $b>0$ for $\omega_s>0$. Besides the condition described in Eq. (\ref{Doublet Yukawa conditions}), the couplings $\lambda_2, \lambda_4$ and $\lambda'_4$ also have to maintain the following constraints among themselves to keep all other mass matrices positive-definite:
\begin{align}
3 \lambda_2 +3 \lambda_4 +4 \lambda'_4 &>0; \nonumber\\
\lambda_2 + \lambda_4 + 4 \lambda'_4 &>0; \nonumber\\
2 \lambda_2 + 3 \lambda_4 + 2 \lambda'_4 &>0; \nonumber\\
 \lambda_2 +  \lambda_4  &>0;\nonumber\\
8( \lambda_2 +  \lambda_4 + 2 \lambda'_4) \sigma^2 +4 \beta \omega_s^2 &>0;\nonumber\\
8( \lambda_2 +  \lambda_4 - 2 \lambda'_4) \sigma^2 + \beta \omega_s^2 &>0.\nonumber
\end{align}

After going through the process described before and keeping track of all the consistency checks and phenomenological constraints, one can produce an ample amount of benchmark points. One can adopt a one-loop RGE evolution while updating the beta functions as one arrives at the threshold of each scalar multiplet. Or one can adopt a two-loop RGE evolution while keeping all the threshold corrections at the corresponding scale. In the next couple of subsections we present our results for each cases.

\subsection*{Benchmark point using one-loop RGE}
For the first case we consider the evolution of gauge couplings using one-loop RGE and include the effect of a scalar multiplet in the beta coefficients at the threshold energy scale corresponding to its mass. The benchmark point we select is given in Table \ref{Table: Sample point 1 loop} and corresponding mass spectrum is given in Table \ref{Table: Sample mass 1 loop}.
\begin{table}[!ht] 
\centering
\begin{center}
\begin{tabular}{c|c||c|c}
 \hline 
 Parameter &Value & Parameter & Value \\ 
 \hline 
$b$ & $ 1.70$ & $a$ & $0.31$\\
$\lambda_2$ & $-0.17$ & $\lambda_0$ & $0.90$  \\ 
$\lambda_4$ & $0.48$ & $\alpha $ & $-0.23$ \\
$\lambda_4'$ & $0.17$ & $\chi_1$ & $0.10$\\
$\beta$ & $ 1.25 \times 10^{-5}$ & $\chi_2$ & $0.12$\\
$\eta_1$ & $-0.002$ & $\chi_3$ & $-0.01$\\
$\eta_2$ & $0.90$ & $c$ & $9.36 \times 10^{15}$ GeV\\
$\chi_4$ & $-0.55$ & $\xi_3$ & $ -3.15\times 10^{14} $ GeV\\
$\chi_5$ & $ 0.32$ & $\chi_6$ & $-2.67 \times 10^{14}$ GeV\\
$\gamma_1$ & $-0.38$ & $ v_s$ & $9.36 \times 10^{10}$ GeV\\
$\gamma_2$ & $0.52$ & $ \sigma$ & $8.65 \times 10^{14}$ GeV\\
$\eta_0$ & $-0.15$ & $ \omega_s$ & $1.38 \times 10^{16}$ GeV\\
 \hline
 \end{tabular}  
 \end{center}
 \caption{Sample parameters and vev's to generate a benchmark point using one-loop RGE. The initial parameter and the vev values were updated through the iteration processes described in the text, and the listed values correspond to the final stable point.}
\label{Table: Sample point 1 loop}
\end{table}

\begin{table}[!ht]
\centering
\begin{center}
\begin{tabular}{c|c||c|c} 
 \hline 
 Multiplet & Mass [GeV] &  Multiplet & Mass [GeV]\\ 
 \hline 
$(1,3,0)$ & $ 2.54 \times 10^{16}$ &$(8,1,0)$ & $ 1.11 \times 10^{12}$\\
\hline
$(3,3,${\footnotesize $-\frac{1}{3}$}$)$ & $2.17 \times 10^{14}$ & $(6,3,${\footnotesize $ +\frac{1}{3}$}$)$ & $2.42\times 10^{14}$  \\
\hline
$(1,1,+2)$ & $2.42 \times 10^{14}$ & $(\overline{3},1,${\footnotesize $+\frac{4}{3}$}$)$ & $2.17 \times 10^{14}$  \\
\hline
$(\overline{6},1,${\footnotesize $-\frac{4}{3}$}$)$ & $ 2.42\times 10^{14}$ & $(\overline{6},1,${\footnotesize $-\frac{1}{3}$}$)$ & $2.17 \times 10^{14}$  \\ 
\hline
 \multirow{2}{*}{$(1,3,-1)$}  & $2.54 \times 10^{16}$ &\multirow{2}{*}{$(\overline{6},1,${\footnotesize $+\frac{2}{3}$}$)$} & $1.13 \times 10^{12}$ \\ 
\hhline{~-~-}
 • & $2.91 \times 10^{14}$ & • & $2.91 \times 10^{14}$ \\ 
\hline
 \multirow{2}{*}{$(3,2,${\footnotesize $+\frac{7}{6}$}$)$}  & $2.47 \times 10^{14}$ &\multirow{2}{*}{$(8,2,${\footnotesize $-\frac{1}{2}$}$)$} & $ 2.47\times 10^{14}$ \\ 
\hhline{~-~-}
 • & $2.22 \times 10^{14}$ & • & $ 2.22\times 10^{14}$ \\
 \hline
 \multirow{2}{*}{$(3,2,${\footnotesize $+\frac{1}{6}$}$)$}  & $4.83 \times 10^{13}$ &\multirow{4}{*}{$(1,2,${\footnotesize $-\frac{1}{2}$}$)$} & $2.23 \times 10^{14}$ \\ 
\hhline{~-~-}
 • & $2.95 \times 10^{14}$ & • & $8.12 \times 10^{13}$ \\
 \hhline{--~-}
 \multirow{5}{*}{$(3,1,${\footnotesize $-\frac{1}{3}$}$)$}  & $2.86 \times 10^{14}$ &• & $1.13 \times 10^{12}$ \\ 
\hhline{~-~-}
 • & $2.37 \times 10^{14}$ & • & $\approx 0$ \\
\hhline{~---}
 • & $8.22 \times 10^{14}$ & \multirow{3}{*}{$(1,1,0)$} & $ 1.66\times 10^{16}$ \\
\hhline{~-~-}
 • & $6.99 \times 10^{13}$ & • & $ 3.90 \times 10^{13}$ \\
\hhline{~-~-}
 • & $1.28 \times 10^{13}$ & • & $ 2.69 \times 10^{10}$ \\
 \hline
 \end{tabular}  
 \end{center}
\caption{Sample scalar mass spectrum corresponding to the benchmark point generated using one-loop RGE. The value of the parameters and vev's used to generate the spectrum is given in Table \ref{Table: Sample point 1 loop}.}
\label{Table: Sample mass 1 loop}
\end{table}
\begin{figure}
\centering
\begin{subfigure}{.8\textwidth}
  \centering
  \includegraphics[width=.95\linewidth]{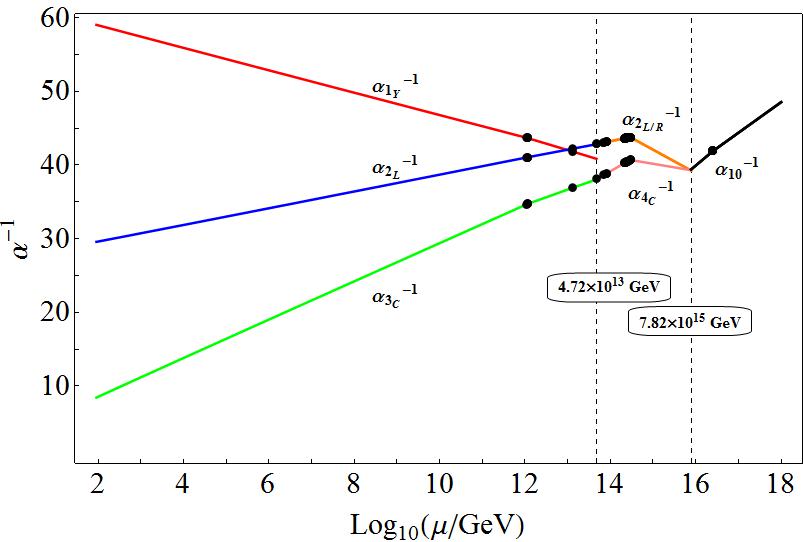}
   \label{fig:RGE one loop - sample point 02}
  \caption{}
\end{subfigure}
\par\bigskip
\begin{subfigure}{.8\textwidth}
  \centering
  \includegraphics[width=0.95\linewidth]{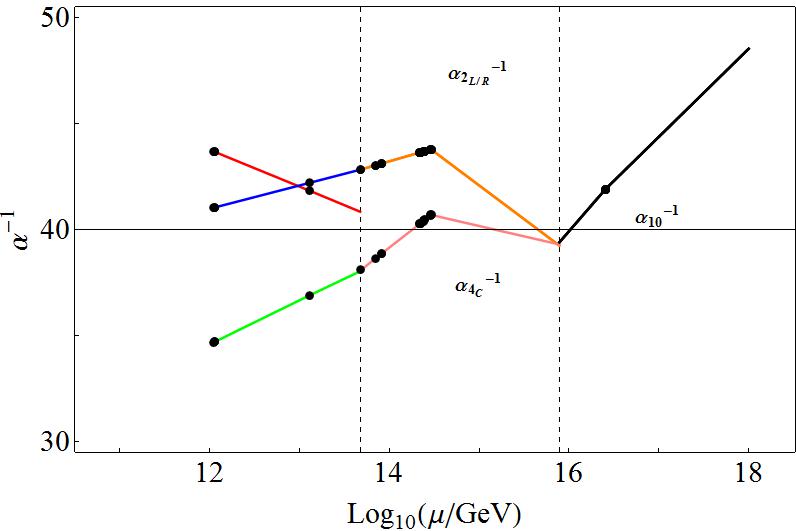}
     \caption{}
  \label{fig:RGE one loop - sample point 02 zoomed}
\end{subfigure}
\caption{(a) Evolution of gauge couplings using one-loop RGE with threshold corrections determined by the scalar mass spectrum given in Table \ref{Table: Sample mass 1 loop}. The unification scale determined here is compatible with the current experimental limit on proton lifetime. The small black circles correspond to the various scalar masses changing the $\beta$ function coefficients and inflicting changes in the slope of the graphs. The vertical dashed lines correspond to gauge boson masses that stay at intermediate scale and unification scale. (b) The region where the scalar bosons  show up has been zoomed.}
\label{fig:RGE one loop - sample point}
\end{figure}

For such a sample point, the RGE evolution produces a unification corresponding to a $(X,Y)$ gauge boson mass compatible with the current proton lifetime. For this benchmark point, the $(X,Y)$ gauge bosons have a mass of $7.82 \times 10^{15}$ GeV. Using Eq. (\ref{eq:proton lifetime formula}) we find out the proton lifetime to be  $5.72 \times 10^{34} \; \text{yrs}$, which is permitted by the current experimental limit, but reachable in the next upgrade of proton decay detectors. Also the vev ratios are capable of reproducing fermion mass spectrum as, $r=69, s= 0.36, r_R \approx 10^{-14}$ as demanded by the fermion mass fitting shown in Ref. \cite{Joshipura:2011nn}.


\subsection*{Benchmark point using two-loop RGE}
After updating the sample point at one-loop level, so that it satisfies all the consistency checks and phenomenological constraints, one can upgrade the procedure to two-loop level, while including all the threshold corrections at the corresponding scales. For such a scenario, the vev's have been chosen to play the role of scales. The sample point for the two-loop case is given in Table \ref{Table: Sample point 2 loop} and the corresponding sample scalar mass spectrum is given in Table \ref{Table: Sample mass 2 loop}.
\begin{table}[!ht] 
\centering
\begin{center}
\begin{tabular}{c|c||c|c}
 \hline 
 Parameter &Value & Parameter & Value \\ 
 \hline 
$b$ & $ 1.70$ & $a$ & $0.31$\\
$\lambda_2$ & $-0.17$ & $\lambda_0$ & $0.90$  \\ 
$\lambda_4$ & $0.49$ & $\alpha $ & $-0.23$ \\
$\lambda_4'$ & $0.17$ & $\chi_1$ & $0.10$\\
$\beta$ & $ 1.25 \times 10^{-5}$ & $\chi_2$ & $0.12$\\
$\eta_1$ & $-0.002$ & $\chi_3$ & $-0.12$\\
$\eta_2$ & $-0.73$ & $c$ & $8.50 \times 10^{15}$ GeV\\
$\chi_4$ & $-0.60$ & $\xi_3$ & $ 1.83\times 10^{15} $ GeV\\
$\chi_5$ & $ 0.32$ & $\chi_6$ & $-2.67 \times 10^{14}$ GeV\\
$\gamma_1$ & $-0.38$ & $ v_s$ & $9.36 \times 10^{10}$ GeV\\
$\gamma_2$ & $0.52$ & $ \sigma$ & $8.56 \times 10^{13}$ GeV\\
$\eta_0$ & $-0.15$ & $ \omega_s$ & $1.25 \times 10^{16}$ GeV\\
 \hline
 \end{tabular}  
 \end{center}
 \caption{Sample parameters and vev's to generate benchmark point using two-loop RGE. The initial parameters and the vev's were updated through the iteration processes described in the text, and the listed values correspond to the final stable point. }
\label{Table: Sample point 2 loop}
\end{table}

\begin{table}[!ht]
\centering
\begin{center}
\begin{tabular}{c|c||c|c} 
 \hline 
 Multiplet & Mass [GeV] &  Multiplet & Mass [GeV]\\ 
 \hline 
$(1,3,0)$ & $ 2.31 \times 10^{16}$ &$(8,1,0)$ & $ 1.11 \times 10^{12}$\\
\hline
$(3,3,${\footnotesize $-\frac{1}{3}$}$)$ & $2.18 \times 10^{14}$ & $(6,3,${\footnotesize $ +\frac{1}{3}$}$)$ & $2.42\times 10^{14}$  \\
\hline
$(1,1,+2)$ & $2.42 \times 10^{14}$ & $(\overline{3},1,${\footnotesize $+\frac{4}{3}$}$)$ & $2.18 \times 10^{14}$  \\
\hline
$(\overline{6},1,${\footnotesize $-\frac{4}{3}$}$)$ & $ 2.42\times 10^{14}$ & $(\overline{6},1,${\footnotesize $-\frac{1}{3}$}$)$ & $2.18 \times 10^{14}$  \\ 
\hline
 \multirow{2}{*}{$(1,3,-1)$}  & $2.31 \times 10^{16}$ &\multirow{2}{*}{$(\overline{6},1,${\footnotesize $+\frac{2}{3}$}$)$} & $1.13 \times 10^{12}$ \\ 
\hhline{~-~-}
 • & $2.92 \times 10^{14}$ & • & $2.92 \times 10^{14}$ \\ 
\hline
 \multirow{2}{*}{$(3,2,${\footnotesize $+\frac{7}{6}$}$)$}  & $2.47 \times 10^{14}$ &\multirow{2}{*}{$(8,2,${\footnotesize $-\frac{1}{2}$}$)$} & $ 2.47\times 10^{14}$ \\ 
\hhline{~-~-}
 • & $2.22 \times 10^{14}$ & • & $ 2.22\times 10^{14}$ \\
 \hline
 \multirow{2}{*}{$(3,2,${\footnotesize $+\frac{1}{6}$}$)$}  & $2.96 \times 10^{14}$ &\multirow{4}{*}{$(1,2,${\footnotesize $-\frac{1}{2}$}$)$} & $2.23 \times 10^{14}$ \\ 
\hhline{~-~-}
 • & $4.38 \times 10^{13}$ & • & $8.12 \times 10^{13}$ \\
 \hhline{--~-}
 \multirow{5}{*}{$(3,1,${\footnotesize $-\frac{1}{3}$}$)$}  & $2.83 \times 10^{14}$ &• & $1.13 \times 10^{12}$ \\ 
\hhline{~-~-}
 • & $2.35 \times 10^{14}$ & • & $\approx 0$ \\
\hhline{~---}
 • & $8.22 \times 10^{13}$ & \multirow{3}{*}{$(1,1,0)$} & $ 1.66\times 10^{16}$ \\
\hhline{~-~-}
 • & $7.06 \times 10^{13}$ & • & $ 3.90\times 10^{13}$ \\
\hhline{~-~-}
 • & $1.28 \times 10^{13}$ & • & $ 2.69\times 10^{10}$ \\
 \hline
 \end{tabular}  
 \end{center}
\caption{Sample scalar mass spectrum corresponding to the benchmark point generated using two-loop RGE. The value of the parameters and vev's used to generate the spectrum is given in Table \ref{Table: Sample point 2 loop}.}
\label{Table: Sample mass 2 loop}
\end{table}
\begin{figure}[!htb]
\centering
\includegraphics[width=.75 \textwidth]{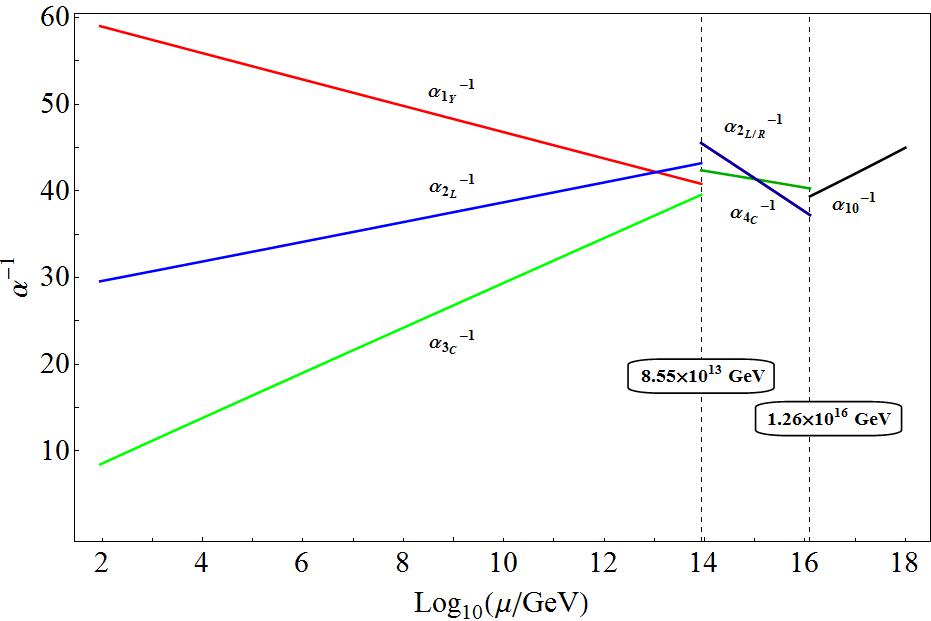}
 \caption{Evolution of gauge couplings using two-loop RGE with threshold corrections. The unification scale determined here is compatible with the current experimental limit on proton lifetime. The discontinuity in the running of the gauge couplings is due to the threshold corrections determined using the scalar mass spectrum given in Table \ref{Table: Sample mass 2 loop}. The vertical dashed lines correspond to the intermediate scale and unification scale.}
\label{fig:Running of Gauge Coupling with Threshold Correction}
\end{figure}

Again for such a sample point, the RGE evolution produces a unification scale corresponding to a $(X,Y)$ gauge boson mass compatible with the current proton lifetime. In this benchmark point, the $(X,Y)$ gauge boson has a mass of $7.11 \times 10^{15}$ GeV. Using Eq. (\ref{eq:proton lifetime formula}) we find out the proton lifetime to be $2.21\times 10^{34} \; \text{yrs}$, which is permitted by the current experimental limit, but reachable in the next upgrade of proton decay detectors. Also the vev ratios are capable of reproducing fermion mass sprectrum as, $r=69, s= 0.36, r_R \approx 10^{-14}$ as demanded by the fermion mass fitting shown in Ref. \cite{Joshipura:2011nn}.

Even though it is desirable to generate Fig. \ref{fig:RGE one loop - sample point} and Fig. \ref{fig:Running of Gauge Coupling with Threshold Correction} from the same sample point, numerically that becomes a difficult task. Even if one starts with the same sample point, due to the updates of parameters and vev coming from the fine-tuning of the doublet mass matrix and iteration process to reduce the error in determining the gauge boson masses (details are described in Sec. \ref{sec:Technical Details}), one ends up with similar, yet not exactly the same sample point. But the uncertainty involved in the process only corresponds to error comparable to higher order loop corrections, and one can claim with enough confidence that final verdict based on such benchmark points is phenomenologically viable in all aspects.
\begin{figure}[!htb]
\centering
\includegraphics[width=.8 \textwidth]{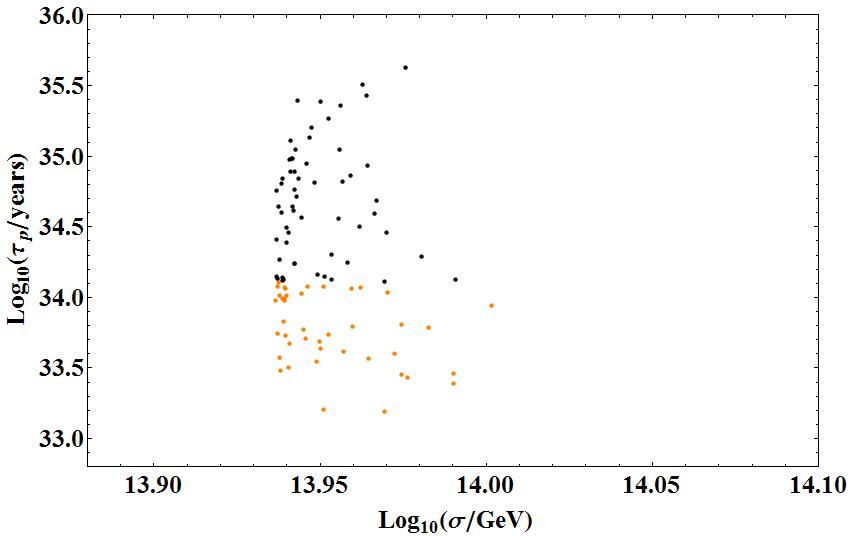}
 \caption{Scatter plot for the proton lifetime $(\tau_p)$ vs.  $\langle 126_H \rangle = \sigma$ generated using one-loop RGE. All the points correspond to proper gauge couplings unification and are compatible with realistic fermion masses and mixings. Only the black points comply with the current experimental limit of proton lifetime.}
\label{fig:scatter plot with threshold}
\end{figure}

One can keep repeating the process and generate multiple points which are phenomenologically viable in all aspects. A scatter plot with such points is shown in the Fig. \ref{fig:scatter plot with threshold}. One major characteristics of the scatter plot with threshold corrections generated using the scalar boson mass spectrum is the distribution of the points, which indicates that the intermediate scale ($\sigma$) does not change much even though the scalar masses are generated with random parameters. This characteristic was missing when the scatter plot was generated with threshold corrections without considering the mass relationship coming into play from scalar mass spectrum. In the absence of such relationships, one can pick the scalar masses completely independently and push the intermediate or unification scale in either direction. But because of the mass relations, due to the fewer number of parameters in $SO(10)$ Lagrangian, one looses such freedom. Selecting one scalar mass in such a way that it will raise the scale fixes mass of another scalar which may tend to lower the scale. Due to the large number of scalar particles in the intermediate scale, the scale tends not to slide much in either direction. But the value of the gauge couplings at the scale do vary from sample to sample.  Similar stationary properties are absent for the case of threshold corrections at the unification scale and one is able to raise the scale high enough to make the proton live long enough to escape the current experimental limit.

Proton lifetime however cannot be raised too much. If one respects the extended survival hypothesis, the upper bound on proton lifetime in this minimal model becomes a few times $10^{35}\; \text{yrs}$. So there is a good possibility of discovering proton decay at Super-Kamiokande  and the next generation experiments.

If one analyzes the scalar mass spectrum carefully, one realizes the fact that all the scalar masses have to remain in the vicinity of intermediate scale and unification scale. One can only introduce extra fine-tuning in the color octet $(8,1,0)$ mass and lower it down without spoiling the whole scenario. This is because its mass is not closely tied to the masses of other scalars and this color octet field does not mediate proton decay. By doing so, one also raises the predicted proton lifetime upto $10^{37}\; \text{yrs}$ which is beyond the reach of next generation proton decay detectors. This is not a likely scenario, since it could mean deviating significantly from the extended survival  or equivalently minimal fine-tuning condition.
\begin{figure}[!htb]
\centering
\includegraphics[width=.8 \textwidth]{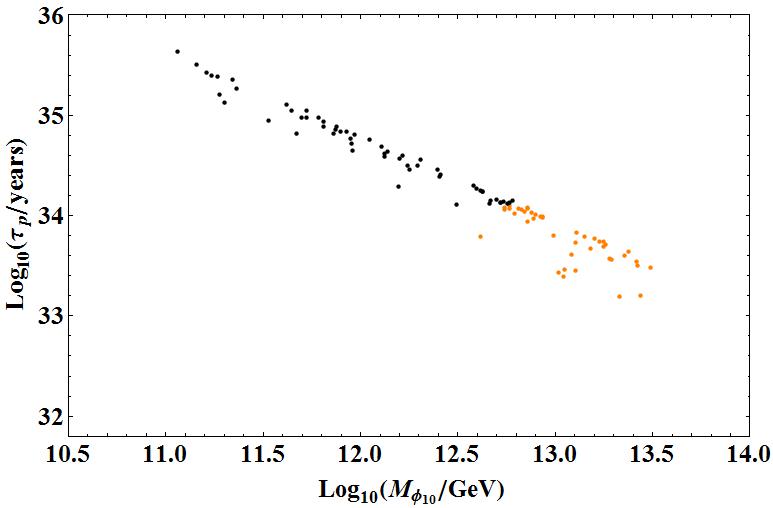}
 \caption{Scatter plot for proton lifetime $(\tau_p)$ as a function of the color octet mass $(M_{\phi_{10}})$ generated using one-loop RGE. All the points are corresponds to proper gauge couplings unification and compatible with realistic fermion masses and mixings. Only the black points comply with the current experimental limit on proton lifetime. Fine-tuning the octet mass to a lower energy scale does not create any internal inconsistency or phenomenological issue. But the extra fine-tuning does mean that one is deviating significantly from the extended survival hypothesis or equivalently minimal fine-tuning condition. That is why we consider highly fine-tuned octet mass which corresponds to a high a proton lifetime as not a likely scenario.}
\label{fig:scattered plot with octet mass}
\end{figure}
\section{Proton decay branching ratios}
\label{sec: Proton branching ratio}
As the color triplets Higgs in the model are always kept heavier than $10^{13}\;\text{GeV}$, the primary source of proton decay is due to $d=6$ gauge boson mediating effective operators which are given by \cite{Nath:2006ut}
\begin{align}
\mathcal{O}^{B-L}_I&=k_1^2 \; \epsilon_{ijk}\; \epsilon_{\alpha\beta}\; \overline{u^C_{ia}}_L\; \gamma^ \mu\; Q_{j\alpha aL}\; \overline{e^C_{b}}_L \;\gamma_\mu \;Q_{k \beta bL};\nonumber \\
\mathcal{O}^{B-L}_{II}&=k_1^2 \; \epsilon_{ijk}\; \epsilon_{\alpha\beta}\; \overline{u^C_{ia}}_L\; \gamma^ \mu\; Q_{j\alpha aL}\; \overline{d^C_{kb}}_L \;\gamma_\mu \;L_{\beta bL};\nonumber \\
\mathcal{O}^{B-L}_{III}&=k_2^2 \; \epsilon_{ijk}\; \epsilon_{\alpha\beta}\; \overline{d^C_{ia}}_L\; \gamma^ \mu\; Q_{j\beta aL}\; \overline{u^C_{kb}}_L \;\gamma_\mu \;L_{\alpha bL};\nonumber \\
\mathcal{O}^{B-L}_{IV}&=k_2^2 \; \epsilon_{ijk}\; \epsilon_{\alpha\beta}\; \overline{d^C_{ia}}_L\; \gamma^ \mu\; Q_{j\beta aL}\; \overline{\nu^C_{b}}_L \;\gamma_\mu \;Q_{k \alpha bL}.
\end{align}
Here, $k_1=\nicefrac{g_{u}}{(\sqrt{2}M_{(X,Y)})}$ and $k_2=\nicefrac{g_{u}}{(\sqrt{2}M_{(X',Y')})}$, $Q_L=(u_L,d_L)$ and $L_L=(\nu_L,e_L)$. The indices $i,j,k$ are color indices, $a,b$ are family indices and $\alpha,\beta$ are $SU(2)_L$ indices. The effective operators in physical basis becomes,
\begin{align}
\mathcal{O}(e_\alpha^C,d_\beta)&= c(e_\alpha^C,d_\beta)\; \epsilon_{ijk}\; \overline{u^C_i}_L\; \gamma^\mu u_{jL} \;\overline{e^C_\alpha}_L\; \gamma_\mu \; d_{k\beta L}; \nonumber \\
\mathcal{O}(e_\alpha,d_\beta^C)&= c(e_\alpha,d_\beta^C)\; \epsilon_{ijk}\; \overline{u^C_i}_L\; \gamma^\mu u_{jL}\; \overline{d^C_{k \beta}}_L\; \gamma_\mu \;e_{\alpha L}; \nonumber 
\end{align}
\begin{align}
\mathcal{O}(\nu_l, d_\alpha,d_\beta^C)&= c(\nu_l, d_\alpha,d_\beta^C)\; \epsilon_{ijk}\; \overline{u^C_i}_L\; \gamma^\mu\; d_{j\alpha L}\; \overline{d^C_{k \beta}}_L\; \gamma_\mu\; \nu_{lL};\nonumber \\
\mathcal{O}(\nu_l^C, d_\alpha,d_\beta^C)&= c(\nu_l^C, d_\alpha,d_\beta^C)\; \epsilon_{ijk}\; \overline{d^C_{i \beta}}_L \; \gamma^\mu\; u_{j L}\; \overline{\nu_l^C}_L \;\gamma_\mu \;d_{k \alpha L};
\end{align}
where
\begin{align}
c(e^C_\alpha,d_\beta) &= k_1^2 \left[ V_1^{11} V_2^{\alpha \beta} +\left( V_1 V_{UD}\right)^{1\beta}\left(V_2 V^\dagger_{UD}\right)^{\alpha 1} \right]; \nonumber\\
c(e_\alpha,d^C_\beta) &= k_1^2 V_1^{11} V_3^{\beta \alpha}+k_2^2 \left(V_4V^\dagger_{UD}\right)^{\beta 1} \left( V_1 V_{UD} V_4^\dagger V_3\right)^{1 \alpha}; \nonumber \\
c(\nu_l,d_\alpha,d_\beta^C)&=k_1^2 \left( V_1 V_{UD}\right)^{1\alpha}\left( V_3 V_{EN}\right)^{\beta l} +k_2^2 V_4^{\beta \alpha} \left( V_1 V_{UD} V_4^\dagger V_3 V_{EN}\right)^{1l};\nonumber\\
c(\nu_l^C,d_\alpha,d_\beta^C)&= k_2^2 \left[\left(V_4 V_{UD}^\dagger \right)^{\beta 1} \left(U_{EN}^\dagger V_2\right)^{l \alpha} + V_4^{\beta \alpha} \left(U_{EN}^\dagger V_2 V_{UD}^\dagger\right)^{l 1}\right]; \alpha=\beta \neq 2.
\end{align}
The mixing matrices are defined as : $V_1=U_C^\dagger U$, $V_2=E_C^\dagger D$, $V_3=D_C^\dagger E$, $V_4=D_C^\dagger D$, $V_{UD}=U^\dagger D$, $V_{EN}=E^\dagger N$ and $U_{EN}=E^\dagger_C N_C$, where U,D,E define the Yukawa coupling diagonalization so that
\begin{align}
U_C^T Y_U U=Y_U^{diag}; \hspace{1cm} & D_C^TY_DD=Y_D^{diag}; \nonumber \\
E_C^TY_EE=Y_E^{diag}; \hspace{1cm} & N^TY_NN=Y_N^{diag}.
\end{align}
For the $SO(10)$ model with symmetric Yukawa couplings, the mixing matrices becomes $U_C=UK_u$, $D_C=DK_d$ and $E_C=EK_e$, where $K_u, K_d$ and $K_e$ are diagonal matrices containing three phases. The proton decay rate into different channels due to the presence of the gauge mediated $d=6$ operators are given by:
\begin{align}
\Gamma(p\rightarrow K^+\overline{\nu})&=\dfrac{\left( m_p^2-m_K^2\right)^2}{8 \pi m_p^3 f_\pi^2} R_L^2 A_S^2 \left| \alpha \right|^2 \sum\limits^3_{i=1} \left| \dfrac{2 m_p}{3m_B}D \; c\left( \nu_i,d,s^C \right)+\left[1+\dfrac{m_p}{3 m_B}\left(D+3F\right)\right] c\left(\nu_i,s,d^C\right)\right|^2; \nonumber \\
\Gamma\left(p \rightarrow \pi^+ \overline{\nu} \right) &=\dfrac{m_p}{8 \pi f_\pi^2} R_L^2 A_S^2  \left| \alpha \right|^2 \left( 1+D+F \right) \sum\limits^3_{i=1} \left|  c \left( \nu_i,d,d^C \right) \right|^2; \nonumber \\
\Gamma\left(p \rightarrow \eta e^+_\beta \right) &=\dfrac{\left( m_p^2-m_\eta^2\right)^2}{48 \pi m_p^3 f_\pi^2} R_L^2 A_S^2 \left| \alpha \right|^2 \left( 1+D-3F \right)^2 \left\lbrace \left| c\left(e_\beta,d^C\right)\right|^2 +\left| c\left(e_\beta^C,d\right)\right|^2 \right\rbrace; \nonumber \\
\Gamma \left( p \rightarrow K^0 e^+_\beta \right) &= \dfrac{\left( m_p^2-m_K^2\right)^2}{8 \pi m_p^3 f_\pi^2} R_L^2 A_S^2 \left| \alpha \right|^2 \left[1+\dfrac{m_p}{m_B} \left(D-F \right)\right]^2 \left\lbrace \left| c\left(e_\beta,s^C\right)\right|^2 +\left| c\left(e_\beta^C,s\right)\right|^2 \right\rbrace; \nonumber \\
\Gamma \left( p \rightarrow \pi^0 e^+_\beta \right) &= \dfrac{m_p}{16 \pi f_\pi^2} R_L^2 A_S^2 \left| \alpha \right|^2 \left(1+D+F \right)^2 \left\lbrace \left| c\left(e_\beta,d^C\right)\right|^2 +\left| c\left(e_\beta^C,d\right)\right|^2 \right\rbrace;
\end{align}
where, $\nu_i=\nu_e,\nu_\mu,\nu_\tau$ and $ e_\beta = e, \mu$. Here $m_B$ is the average baryon mass satisfying $m_B\approx m_\Sigma \approx \Lambda$. As the current (and most probably next generation) proton decay detectors are insensitive to the flavor of the neutrinos, proton decay rates are calculated by summing over all the flavors. For similar reason the chirality of the charged lepton is also summed over. Here, $A_S \approx 2$ is the average of the left-handed and right-handed short range renormalization factor. 

Now if we consider the fermion masses and mixings given in the ref \cite{Joshipura:2011nn}, using the vev ratio parameter values $r=69$ and $s=0.36$, the proton decay branching ratio due to gauge mediated $d=6$ operator are given in the Table \ref{Table: branching ratio}.
\begin{table}[!ht]
\centering
\begin{center}
\begin{tabular}{cc} 
 \hline 
Process & Branching ratio\\
 \hline 
 \hline
 $p \rightarrow \pi^0 e^+$ & $\approx 47 \%$\\
$ p \rightarrow \pi^0 \mu^+$& $\approx 1.00 \% $ \\
$p \rightarrow \eta^0 e^+$& $\approx  0.20 \% $ \\
$p \rightarrow \eta^0 \mu^+$& $\approx 0.004 \% $ \\
$p \rightarrow K^0 e^+$ & $\approx 0.16 \% $ \\
$p \rightarrow K^0 \mu^+$ & $\approx 3.62 \% $ \\
$p \rightarrow \pi^+ \overline{\nu}$ & $ \approx 48 \%$ \\
$p \rightarrow K^+ \overline{\nu}$& $\approx  0.22 \% $\\
 \hline
 \end{tabular}  
 \end{center}
\caption{The branching ratio of proton decay by gauge mediated $d=6$ operator.}
\label{Table: branching ratio}
\end{table}

\begin{figure}[!htb]
\centering
\includegraphics[width=.8 \textwidth]{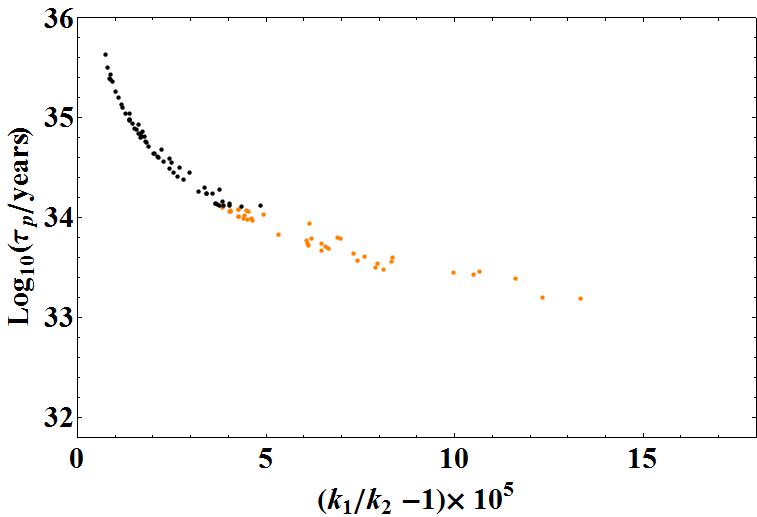}
 \caption{Proton Lifetime$(\tau_p)$ vs ratio of the $\left( \nicefrac{k_1}{k_2} -1 \right) \times 10^5$. The plot indicates that the ratio of $\nicefrac{k_1}{k_2}= \nicefrac{M_{(X',Y')}}{M_{(X,Y)}}$ varies less than $0.02\%$ $(0.005\%)$ over the whole (phenomenologically viable) parameter space.}
\label{fig:ratio of unification gauge boson}
\end{figure}

These branching ratios mainly depend on the ratio of the leptoquark gauge bosons $\nicefrac{k_1}{k_2}=\nicefrac{M_{(X',Y')}}{M_{(X,Y)}}$. From the gauge boson mass spectrum and the scatter plot (Fig. \ref{fig:ratio of unification gauge boson}), it is clear that in this $SO(10)$ model, the branching ratios will not vary much within the phenomenologically viable parameter space. We see that the dominant modes are $ p \rightarrow e^+ \pi^0 $ and $p \rightarrow \overline{\nu} \pi^+$, with roughly equal rates.

The proton decay branching ratios given in Table \ref{Table: branching ratio} is quite similar to the one given in Ref.~\cite{Bajc:2008dc} for the case of minimal $SO(10)$ with split supersymmetry. This is mainly due to the fact that the Yukawa sector is essentially the same (upto renormalization effects) and since  $M_{(X,Y)} \backsimeq M_{(X',Y')}$ was assumed in Ref.~\cite{Bajc:2008dc}.

\section{Axions as Dark Matter}
\label{sec:Axions as Dark Matter}
Introducing a Peccei-Quinn(PQ)-symmetry \cite{Peccei:1977hh} in non-supersymmetric $SO(10)$ GUTs provides the perfect framework for the axionic dark matter which simultaneously solves the strong CP problem. The PQ-symmetry affected the Higgs potential (by removing terms like $126^4$) and also made the Yukawa sector realistic and predictive. Yet, the main contribution of the PQ-symmetry is to provide axion as dark matter candidate which can explain the entire dark matter abundance in the universe while also solving the strong CP problem.

The axion in the model is of DFSZ type \cite{Dine:1981rt, Zhitnitsky:1980tq}. While the original DFSZ axion was mainly composed of a complex singlet field with an admixture of one up-type Higgs doublet and one down-type Higgs doublet, the axion in this model is mainly composed of the complex singlet field ($S_H$) with the admixture of two up-type Higgs doublets ($H_u$ from $126_H$ and $h_u$ from $10_H$) and two down type Higgs doublets ($H_d$ from $126_H$ and $h_d$ from $10_H$).

In the model, the PQ-symmetry is broken by the vev of the singlet $S_H$, and the scale is quiet independent of the intermediate (Pati-Salam) scale and unification scale. Even though the choice of $v_s$ is mainly guided by the axion phenomenology, in the numerical analysis of the sample points we found out that the PQ-breaking scale stays around $\left( 5 \times 10^{10} - 1 \times 10^{12}\right) \;\text{GeV}$ without any extra fine-tuning. In that case, the axion mass can be computed using 
\begin{equation}
m_a = \dfrac{z^{\nicefrac{1}{2}}}{1+z} \dfrac{ f_\pi m_\pi}{f_a}
\end{equation}
where $z= \nicefrac{m_u}{m_d}$ and $f_a$ is the axion decay constant. For the numerical analysis we took $m_\pi=135 \; \text{MeV}$ and $f_\pi \approx 130.7 \; \text{MeV}$ and kept the range of $z = 0.35-0.60$ \cite{Beringer:1900zz}. Then for $f_a=v_s$, we get $m_a ~ (8 - 175)\; \text{$\mu$eV}$, which is compatible with both the laboratory experimental limit and astrophysical bounds.
\subsection*{PQ-symmetry breaking before or during inflation}
The cosmic mass density of axion field today is  \cite{Sikivie:2006ni}
\begin{equation}
\Omega_a h^2 \approx 0.7 \left( \dfrac{f_a}{10^{12}\; GeV} \right)^{\nicefrac{7}{6}} \left(\dfrac{\overline{\Theta}_i}{\pi} \right)^2,
\end{equation}
where $h$ is the present-day Hubble expansion parameter and $-\pi \geq \overline{\Theta}_i \geq \pi$ is the initial ``misalignment angle". If the PQ-symmetry is broken before or during inflation, inflation expands a domain with some value of $\overline{\Theta}_i$ to a size larger than the present universe. In that case, $\overline{\Theta}_i$ can take any value and naturally should not be fine-tuned. Using the experimental limit $\Omega_a h^2 = 0.1199 \pm 0.0027$ \cite{Ade:2015xua}, we can scan the parameter space in the ${\alpha, f_a}$ basis.
\begin{figure}[!htb]
\centering
\includegraphics[width=.8 \textwidth]{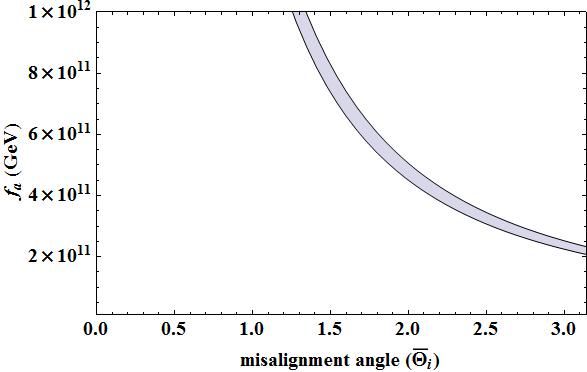}
 \caption{The region in the parameter space where cold ADM saturates the dark matter abundance.}
\label{fig:ADM parameter space}
\end{figure}

As the model allows the axion decay constant as low as $5 \times 10^{10} \; \text{GeV}$ and as high as $10^{12} \; \text{GeV}$, the misalignment angle can take any value beyond 1.26, ie $1.26<\left| \overline{\Theta}_i \right| < \pi$. We also see that, for $f_a$ smaller than $2.33 \times 10^{11}\; \text{GeV}$, axionic dark matter fails to explain the entirety of dark matter abundance. 

From the recent Planck data, we find that if we assume that the PQ symmetry is broken during inflation and it is not restored by the quantum fluctuation of the inflation nor by thermal fluctuation in the case of a very efficient reheating stage and all of cold dark matter (CDM) consists of axions produced by the misalignment angle mechanism, the upper bound on the energy scale of inflation ($H_{inf}$) becomes\cite{Planck:2013jfk}:
\begin{equation}
H_{inf}\leq 0.87 \times 10^7 \; GeV \left(\dfrac{f_a}{10^{11}\; GeV} \right)^{0.408}.
\end{equation}
This is due to the fact that the axion which already exists during inflation obtains large quantum fluctuations and produces isocurvature density perturbations which are stringently constrained by CMB observation. So, for low axion decay constant, we end up having an upper bound on the energy scale of inflation as low as $~10^7 \; \text{GeV}$.
\subsection*{PQ-symmetry breaking after inflation}
Cosmological consequences of axions are different if the PQ-symmetry is broken after inflation. Unlike the previous case, universe does not settle into the same minimum when the axion acquires its mass at the QCD scale and ends up forming topological defects \cite{Kawasaki:2013ae}. Now, the misalignment axion cold dark matter energy density is given by \cite{DiValentino:2014zna}
\begin{equation}
\Omega_{a,mis}h^2= 2.07 \left( \dfrac{f_a}{10^{12}\; GeV} \right)^{\nicefrac{7}{6}}
\end{equation}
while
\begin{equation}
\Omega_a h^2= 2.07(1+\alpha_{dec}) \left( \dfrac{f_a}{10^{12}\; GeV} \right)^{\nicefrac{7}{6}}
\end{equation}
where $\alpha_{dec}=0.164$ corresponds to the factor introduced due to the decay of topological defects like axionic strings \cite{DiValentino:2014zna}. Under such consideration, one finds that the Planck data corresponds to a axion mass $m_a \approx 80\; \text{$\mu$eV}$ and axion decay constant $f_a \approx 8 \times 10^{10}\; \text{GeV}$ \cite{DiValentino:2014zna}, which are perfectly admissible in the GUT under scrutiny.

\section{Conclusion}
\label{sec:Conclusion}
The Standard Model emerged in the early seventies and since then it has been weathered by all sorts of experiments at various laboratories and colliders. Until now, it has given the best description of nature. Recent discoveries about dark matter, neutrino masses and mixings and old questions like charge quantization and baryogenesis demand physics beyond the SM, yet LHC data upto now has failed to provide any glimpse of such new physics. In the realm of unification models, the supersymmetric $SO(10)$ GUTs have been studied in depth in the past decades \cite{Babu:2010ej}. The crucial point about $SO(10)$ GUTs is that if we change our current attitude about fine-tuning, yet keep it at the minimal level by adopting philosophy like extended survival hypothesis, we realize that even without supersymmetry, $SO(10)$ symmetry has the potential to be the gauge symmetry of nature on its own right, atleast upto the GUT scale $(\sim 10^{16}\; \text{GeV)}$. The absence of low energy supersymmetry might be the reality of our universe, taking away primary motivation to introduce supersymmetry. Thus it becomes mandatory to revisit the non-supersymmetric version of $SO(10)$ GUTs with a more open attitude. 

The purpose of the paper was to search for the minimal non-supersymmetric $SO(10)$ grand unified model which can withstand the pressure of all the phenomenological constraints. Our aim was to address all possible issues (except gravity) either explicitly or by showing that the model has enough flexibility to accommodate the phenomena. We acknowledge that the minimality is not a universal and uniquely defined concept. In this work, the philosophy of minimality was applied in the choice Higgs representation and that resulted in a breaking pattern with minimal number of intermediate scale (namely one) making the model truly minimal and predictive. 

Such a minimal model ended up relying on threshold corrections to escape from the wrath of experimental bounds on proton lifetime. The issue of threshold corrections deserves particular attention here. On one hand, one should not discard a model without taking into account the threshold corrections, on the other hand, one should not expect that threshold corrections can rescue any model before performing detailed calculation.

The non-SUSY $SO(10)$ GUT presented here managed to unify the gauge couplings at a scale high enough to comply with the current experimental bound of proton lifetime. The Yukawa sector of the model provided a realistic description of fermion masses and mixings. The PQ-phase transition introduced axion as the dark matter candidate that can explain the dark matter abundance in the universe, while also solving the strong CP problem. Leptogenesis finds a natural place in $SO(10)$ with seesaw mechanism and the Yukawa sector of the model has the potential to procure the right amount. Physics of inflation may reside outside the scope of the model or within the model where one (or more) SM singlets already present may provide the necessary ingredients. 

One should emphasis the claim that the SM spectrum is completed by the recently discovered light Higgs and LHC should fail to find any other new physics, as the next scale of physics lies at the energy scale of $10^{10}\; \text{GeV}$. Before getting demoralized one also needs to realize that the model generally predicts a proton lifetime less than a few times $10^{35} \; \text{yrs}$. So Super-Kamiokande or next generation proton decay detectors and axion search experiments has the potential to discover the essential phenomenological proof of the model.
\section*{Acknowledgments}
The authors would like to thank the organizers of CETUP* 2015 for hospitality and partial support during the 2015 Summer Program at Lead, South Dakota where the work was completed. The authors would also like to thank the  participants of CETUP* 2015 for helpful discussions and comments. This work has been supported in part by the U. S. Department of Energy Grant No. de-sc0010108.

\end{document}